\documentclass{aa}  

\usepackage{float}
\usepackage{graphicx}
\usepackage{txfonts}
\usepackage{placeins}           
\usepackage{amsmath}	        
\usepackage{advdate}
\usepackage[]{hyperref}
\usepackage{xcolor}
\hypersetup{colorlinks,
            linkcolor=red,
            citecolor=blue,
            filecolor=cyan,
            urlcolor=magenta}
\usepackage{orcidlink}
\newcommand{\Msun}{\mbox{$M_{\odot}$}}
\newcommand{\AU}{\mbox{$\mathrm{au}$}}
\newcommand{\gcep}{\mbox{$\gamma$~Cephei}}           
\newcommand{\MA}{\mbox{$M_{\mathrm{A}}$}}            
\newcommand{\MB}{\mbox{$M_{\mathrm{B}}$}}            
\newcommand{\Mp}{\mbox{$M_{p}$}}                     
\newcommand{\Mjup}{\mbox{$M_{\mathrm{J}}$}}          
\newcommand{\St}{\mbox{$\mathrm{St}$}}               
\newcommand{\cisec}[1]{Section~\ref{#1}}             
\newcommand{\cifig}[1]{Figure~\ref{#1}}              
\newcommand{\cieq}[1]{Eq.~(\ref{#1})}           

\graphicspath{{./}{./figures/}} 

\begin{document}

   \title{Gas and dust dynamics in $\gamma$ Cephei-type disks}

   \author{
           Francesco Marzari\inst{1}\orcidlink{0000-0003-0724-9987}
        \and
           Gennaro D'Angelo\inst{2,}\thanks{Corresponding author: \texttt{gennaro@lanl.gov}}\orcidlink{0000-0002-2064-0801}
        }

   \institute{
   Department of Physics and Astronomy, University of Padova, via Marzolo 8, I-35131, Padova, Italy\\
             \email{francesco.marzari@pd.infn.it}
             \and
   Theoretical Division, Los Alamos National Laboratory, Los Alamos, NM 87545, USA\\
             \email{gennaro@lanl.gov}
             }

   \date{Received \today}

  \abstract
   {Giant planets are observed orbiting the primary stars of close binary
   systems. Such planets may have formed in compact circumprimary 
   disks, which once surrounded these stars, under conditions much 
   different than those encountered around single stars.
   }
   {In order to quantify the effects of the strong gravitational 
   perturbations exerted on circumprimary disk material, 
   the three-dimensional (3D) dynamics of gas and dust in orbit around
   the primary star of a compact and eccentric binary system was modeled
   by applying the stellar and orbital parameters of \gcep,
   a well-known system that can be representative of a class
   of close binaries.
   }
   {Circumprimary gas was approximated as an Eulerian viscous and 
   compressible fluid and modeled by means of 3D hydrodynamical 
   simulations, assuming locally isothermal conditions in
   the medium around the primary star.
   Dust grains were modeled as Lagrangean particles, subjected 
   to gravity and aerodynamic drag forces. 
   Models that include a giant planet were also considered.}
   {Models indicate that spiral density waves excited around
   pericenter passage propagate toward the inner boundary of the disk,
   through at least a few pressure scale-heights from the mid-plane,
   inducing radial and vertical mixing in the gas.
   However, perturbations imparted to gas, both
   in terms of eccentricity and precession, are far weaker than
   previously estimated by two-dimensional (2D) simulations.
   Models predict small eccentricities, $\lesssim 0.03$, and 
   slow retrograde precession. The addition of a giant planet
   does not change the low eccentricity state of the disk.
   The parameters applied to the disk would lead to the formation
   of a massive planet, many times the mass of Jupiter,
   in agreement with some observations.
   Micron to mm-size dust grains are well coupled to the gas,
   resulting in similar dynamics and statistically similar
   distributions of orbital elements.
   The planet only affects the dust distributions locally.
   In agreement with outcomes of recent 2D models, the lifetime
   of an isolated circumprimary disk would be brief, $\sim 10^{5}$
   years, because of its compact nature, requiring a long-term
   external supply of mass to allow for the in situ
   formation of a giant planet.
   }
   {}

   \keywords{Accretion, accretion disks --
             Methods: numerical --
             Protoplanetary disks --
             Stars: binaries: close
               }

   \titlerunning{Gas and dust dynamics in $\gamma$ Cephei-type disks}
   \authorrunning{D'Angelo and Marzari}

   \maketitle

\section{Introduction}
\label{sec:intro}
\nolinenumbers
\defcitealias{marzari2025}{MD25}

$\gamma$~Cephei is a $3\,\mathrm{Gyr}$-old, close-binary system 
located approximately $14$~parsecs away \citep{baines2018}. 
The presence of a low-mass stellar companion, \gcep~B, was first 
reported by \citet{campbell1988} and first imaged by \citet{neuhauser2007}.
The primary star, \gcep~A, hosts a substellar-mass companion within
$\approx 2\,\AU$ \citep{hatzes2003}.
If this object formed around the primary star, it did so in a 
highly perturbed environment, considering that it orbits at around 
$1/6$ of the binary's pericenter distance.
Importantly, \gcep\ is not peculiar in this regard, as planetary-mass 
companions observed in compact and eccentric binaries are not rare.
Indeed, \gcep\ can be considered as a representative of a class of 
binary systems.

In fact, according to the catalog of exoplanets in binary 
systems\footnote{\url{https://adg.univie.ac.at/schwarz/multiple.html}}
maintained by \cite{schwarz2016},
there are at least $20$ systems, with separations between $10$ and 
$\approx 50\,\AU$, which host a planet more massive than $0.1$ Jupiter
masses ($\Mjup$).
Surprisingly, $60$\% of these binaries are eccentric, with orbital
eccentricities in the range from $\approx 0.1$ to $\approx 0.5$. 
Since tidal interactions between orbiting material and the stars
are expected to truncate the circumprimary disk at $\approx 25$\% of the 
binary's semi-major axis, 
these systems would be surrounded by compact disks extending from a few
to $\approx 12\,\AU$ from the primary.
Therefore, in these systems, planet formation occurs (if it does)
in a very small disk.

An initial assessment on the conditions required to form a massive planet
around \gcep~A indicated the need for a relatively large gas
density in the circumprimary disk \citep{thebault2004}, a condition 
required by the small disk size. Such an outcome may be aided 
by tidal confinement driven by the orbital eccentricity of the binary.
However, in these situations, additional complications may affect formation
by core accretion. A short list includes:
\textit{a}) the circumprimary disk lifetime, which must exceed $\sim 10^{6}$ 
years to allow for the growth of a giant planet;
\textit{b}) the availability and supply of solid material, which must be
sufficient to form the initial core of the planet;
\textit{c}) solid dynamics, which must be conducive to accumulation, rather
than fragmentation, an outcome affected by the highly perturbed gas dynamics.
This paper looks into the details of the circumprimary disk
evolution to provide constraints on both gas and dust dynamics.
We do so by performing three-dimensional (3D) simulations of the evolution
of the gas and dust in a disk orbiting the primary of
a $\gamma$ Cephei-type system. We consider disk configurations both
before and after the possible formation of a giant planet. In particular,
we focus on disk eccentricity, which appears to be significantly smaller
than that estimated by previous two-dimensional (2D) simulations.
In addition, the precession rate of the disk is much slower and both
these results may have profound implications on the evolution
of planetesimals in the disk \citep{beauge2010,silsbee2015}.
Mass transport through the disk, and toward the primary star, is 
enhanced by a factor of a few compared to the expectations of 
unperturbed, steady-state disks around single stars, as a result
of its compact nature and inward propagation of
spiral waves excited around pericenter passage.

When a massive planet is included in the circumprimary disk,
a deep gap forms around the planet's orbit and dust grains accumulate
outside the gap.
We calculate the migration rate of the planet and estimate its
mass accretion rate. Under the assumption of sustained supply
of gas, from an external source, the planet may easily reach
a mass several to many times the mass of Jupiter.
This prediction is in agreement with some observational estimates
\citep{benedict2018}

It was suggested by previous work on compact binaries \citep[][]{picogna2013}
that hydraulic jumps at the locations of spiral density waves might excite
the distribution of the dust orbital inclination, possibly slowing
down the process of sedimentation toward the disk mid-plane and
therefore hindering or preventing streaming instability from forming
planetesimals. In these calculations, we consider a number of particles
large enough, and with different sizes, to test the effects of
the perturbations caused by spiral waves on the orbital evolution of dust.

The results of the models presented herein can be used to characterize
gas and dust dynamics around the primaries of binary systems with masses
comparable to \gcep~A and B, and with similar orbital separations and
eccentricities.
Applicability of these models is obviously constrained by the physical
description of the system adopted in the simulations.
The applied methods are described in \cisec{sec:NM} and the physical
specifications are given in \cisec{sec:PP}. The results of 
the calculations are discussed in \cisec{sec:GD} and \ref{sec:DD}.
We present our conclusions in \cisec{sec:CC}.

\section{Numerical methods}
\label{sec:NM}

The gas surrounding the primary star was approximated to a viscous and 
compressible fluid, through the Navier-Stokes equations, which were solved
in three dimensions by means of a finite-difference algorithm,
second-order accurate in space and time
\citep[see][and references therein]{gennaro2013}. 
Two dimensional versions of the 3D models were also simulated 
to highlight major differences.
In order to mitigate restrictions on the calculation time-step, imposed 
by the Courant–Friedrichs–Lewy condition, orbital advection 
\citep{masset2000} was applied as described in \citet[][]{gennaro2012}.

The 3D disk domain was rendered and discretized in spherical coordinates 
$\{r, \theta, \phi\}$, with origin on the primary star ($\{r, \phi\}$
coordinates were used in 2D; note that we use the same notation for
the polar radius in 3D and the cylindrical radius in 2D).
The domain extends $2\pi$ in azimuth ($\phi$) around the primary and 
is symmetric with respect to the mid-plane ($\theta=\pi/2$). The gravity 
field is given through the potential
\begin{equation}
 \Phi=\Phi_{\mathrm{A}}+\Phi_{\mathrm{B}}+\Phi_{p}+\Phi_{\mathrm{I}},
 \label{eq:PHI}
\end{equation}
where $\Phi_{\mathrm{A}}$ and $\Phi_{\mathrm{B}}$ are the potentials of
the primary and secondary star, respectively, and $\Phi_{p}$ is 
the potential
generated by planet (when applicable). The last term on the right-hand 
side is the indirect potential arising from non-inertial forces generated
by the motion of the origin of the reference frame,
\begin{equation}
 \Phi_{\mathrm{I}}=\frac{G\MB}
            {r^{3}_{\mathrm{B}}}\mathbf{r}\cdot\mathbf{r}_{\mathrm{B}}%
                  +\frac{G\Mp}{r^{3}_{p}}\mathbf{r}\cdot\mathbf{r}_{p}.
 \label{eq:PHII}
\end{equation}
The planet potential, $\Phi_{p}$, was smoothed over a length 
$R_{\mathrm{H}}/5$, where $R_{\mathrm{H}}$ is the planet's
Hill radius (no softening is applied to the stellar potentials).
Gas self-gravity was neglected.
Note that, although the choice of the reference frame is arbitrary,
a reference system with origin on the primary is best suited to describe
the dynamics of circumprimary disk.
In simulations that include the planetary body, the orbits were integrated
by means of a variable order and adaptive step-size Gragg-Bulirsch-Stoer
extrapolation algorithm \citep{hairer1993}, requiring convergence at
machine precision. 

We used two grids with $1142\times 22\times 630$ and
$1902\times 22\times 1048$ grid elements, respectively.
Some models including a giant planet were also computed on a
$1142\times 22\times 1260$ grid.
Models in two dimensions have the same grid structures in the radial 
and azimuthal directions.
Outflow boundary conditions were applied at the outer radial boundary and
reflective boundary conditions were applied at the inner radial boundary
and at the disk surface.
After models achieve equilibrium states, outflow boundary conditions
are also applied at the radial inner boundary so that gas can flow 
toward the primary.
Boundary conditions applied herein do not allow for possible delivery 
of material to the circumprimary disk from a circumbinary disk 
(not modeled in these calculations).
At the level of gas viscosity and pressure scale-height applied in
the models, the tidal barrier maintained by the stars is strong enough
that the supply of both gas and dust to the circumprimary disk would be 
negligible during the simulated timescales.
In fact, \citet[hereafter, \citetalias{marzari2025}]{marzari2025} showed that
a circumbinary disk with comparable gas parameters would supply gas to
the circumprimary disk at a rate $\sim 10^{-8}\,\Msun\,\mathrm{yr}^{-1}$,
adding a trivial amount of mass during the simulations.

Solids were modeled as Lagrangian (collision-less) particles, interacting 
through gravity with the stars and through drag with the gas. Particle 
thermodynamics was computed and evolved as described in 
\citet[][]{gennaro2015}.
The initial distribution of silicate dust comprises $20\,000$ particles,
distributed in equal numbers into four size bins of radius
$R_{s}=1\,\mu\mathrm{m}$,
$10\,\mu\mathrm{m}$, $100\,\mu\mathrm{m}$, and $1\,\mathrm{mm}$.
Since gas temperatures around the primary may significantly exceed
ice evaporation temperatures \citep[see, e.g.,][]{picogna2013,jordan2021},
we modeled silicate grains with a material density 
$\varrho_{s} = 2.648\,\mathrm{g\,cm^{-3}}$.
Solids that move past the radial and vertical boundaries become inactive.
Solids that cross the disk mid-plane are reflected, mimicking dust motion
across the mid-plane.

\section{Physical and numerical parameters}
\label{sec:PP}

\begin{table}[]
\caption{System parameters \label{table:sum}}
\centering
\begin{tabular}{ccc}
\hline\hline
Quantity & Symbol & Value\\
\hline
Primary mass   & \MA   & $1.4\,\Msun$\tablefootmark{a}\\
Secondary mass & \MB   & $0.4\,\Msun$\tablefootmark{a}\\
Binary orbital separation   & $a$ & $20\,\AU$\tablefootmark{b}\\
Binary orbital eccentricity & $e$ & $0.4$\tablefootmark{b}\\
Planet mass    & \Mp   & $1.85\,\Mjup$\tablefootmark{c}\\
Planet orbital separation   & $a_{p}$ & $2.05\,\AU$\tablefootmark{c}\\
Planet orbital eccentricity & $e_{p}$ & $0.049$\tablefootmark{c}\\
Circumprimary disk aspect ratio & $H/r$ & $0.05$\\
Gas turbulence parameter & $\alpha$   & $0.001$\\
Dust material density    & $\varrho_{s}$ & $2.648\,\mathrm{g\,cm^{-3}}$\\
\hline
\end{tabular}
\tablefoot{
\tablefoottext{a}{\citet{neuhauser2007};}
\tablefoottext{b}{\citet{mugrauer2022};}
\tablefoottext{c}{\citet{endl2011}}
}
\end{table}

Stellar and orbital parameters representative of \gcep\ are used to represent
a class of binary systems.
Therefore, since we intend the models to be broadly applicable to close 
and eccentric binaries, the precise values of these parameters are not important.
In fact, there are still inconsistencies in the stellar masses reported
in the literature. Here we choose the values derived by  \citet{neuhauser2007},
$M_\mathrm{A}=1.4\,\Msun$ \citep[see also][]{baines2018} and
$M_\mathrm{B}=0.4\,\Msun$. These values are close to those derived more 
recently by \citet{mugrauer2022}, $M_\mathrm{A}=1.3\,\Msun$ and
$M_\mathrm{B}=0.38\,\Msun$. 
\citeauthor{neuhauser2007} values provide a secondary to primary mass ratio
of $q=0.286$, quite close to the mass ratio resulting from the estimates
of \citeauthor{mugrauer2022}, $q=0.292$.
We note that \citet{torres2007} estimated somewhat lower masses,
$M_\mathrm{A}=1.2\,\Msun$ and $M_\mathrm{B}=0.36\,\Msun$, which nonetheless
result in a similar mass ratio, $q=0.3$.
The binary separation is set to $a=20\,\AU$ and its orbital eccentricity 
to $e=0.4$ \citep{endl2011,mugrauer2022}.
With these parameters, the binary period is $T=66.667$ years.

The primary star, \gcep~A, has a substellar-mass companion whose minimum mass
was estimated at $1.85$ Jupiter masses \citep{endl2011}. The actual mass 
of this object may in fact be much larger. Analysis of astrometry measurements
acquired by the Hubble Space Telescope resulted in a value of $9.4$ 
Jupiter masses \citep{benedict2018}. The model planet, however, has to be
smaller than its final mass since it is still surrounded by gas and, thus,
would be actively accreting \citep[e.g.,][]{bodenheimer2013}.
Lacking firm constraints, we use the parameters derived by \citet{endl2011}:
$\Mp=1.85\,\Mjup$, $a_{p}=2.05\,\AU$ and $e_{p}=0.049$.
The planet's orbit is coplanar with the binary orbit, and both lie
in the disk mid-plane.

The primary disk is initialized with a gas surface density $\Sigma\propto 1/r$,
where $r$ indicates the distance from the primary in both 2D and 3D geometry,
and a relative pressure scale-height  $H/r=0.05$. 
Models apply a local-isothermal equation of state
\begin{equation}
   P=c^{2}_{s} \rho,
   \label{eq:P}
\end{equation}
where $\rho$ is the gas density and $c_{s}$ the local-isothermal sound speed 
(the gas temperature around the primary is $\propto 1/r$). 
Because of \cieq{eq:P} and the neglect of disk self-gravity, gas dynamics
can be re-scaled by the initial density. We use \AU\ as units of length
and $\MA/\AU^2$ as units of surface density.
Viscous stresses in the gas are implemented through a time-constant
kinematic viscosity $\nu=\alpha H^{2}\Omega_{\mathrm{K}}$, where 
$\Omega_{\mathrm{K}}$ is the local Keplerian frequency around the primary
and $\alpha=0.001$.
No artificial viscosity \citep[e.g.,][]{stone1992a} is used in these 
calculations.
A summary of some of the main parameters adopted in the models is reported
in Table~\ref{table:sum}.

According to theoretical expectations based on a comparison between
tidal torques and viscous stresses at Lindblad resonances, for the stellar 
and disk parameters applied herein ($\MB/\MA$, $e$, $\alpha$, and $H/r$),
tidal truncation of the circumprimary disk is expected between the inner 
1:9 and 1:8 commensurability of the binary, around 
$r/a\approx 0.23$--$0.25$ \citep{pawel1994}.
Since these models are intended to describe only the circumprimary disk
and considering the above requirement on tidal truncation, the disk domain
extends from $r=0.005\,a$ to $0.575\,a$ ($0.1$--$11.5\,\AU$) around the primary, 
within the pericenter of the secondary. In 3D, the disk extends about $3.5$ 
pressure scale-heights above the mid-plane.

\section{Gas dynamics}
\label{sec:GD}

This section presents results on the dynamics of the gas in the circumprimary
disk, separately introducing models without a giant planet and models that
include a giant planet. Except for the different gravitational potential,
\cieq{eq:PHI} and (\ref{eq:PHII}), these models use the same physical and
numerical parameters.
Models including a planet were only simulated in three dimensions.

\subsection{Circumprimary disk without a giant planet}
\label{sec:GDnop}

\begin{figure*}
\centering%
\resizebox{\linewidth}{!}{\includegraphics[clip]{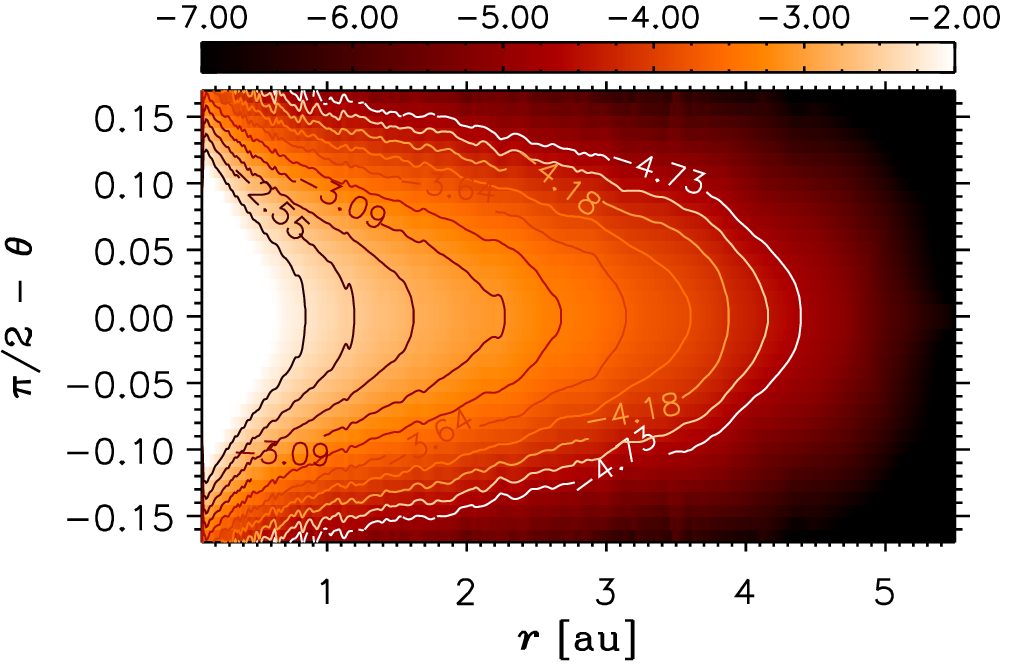}%
                          \includegraphics[clip]{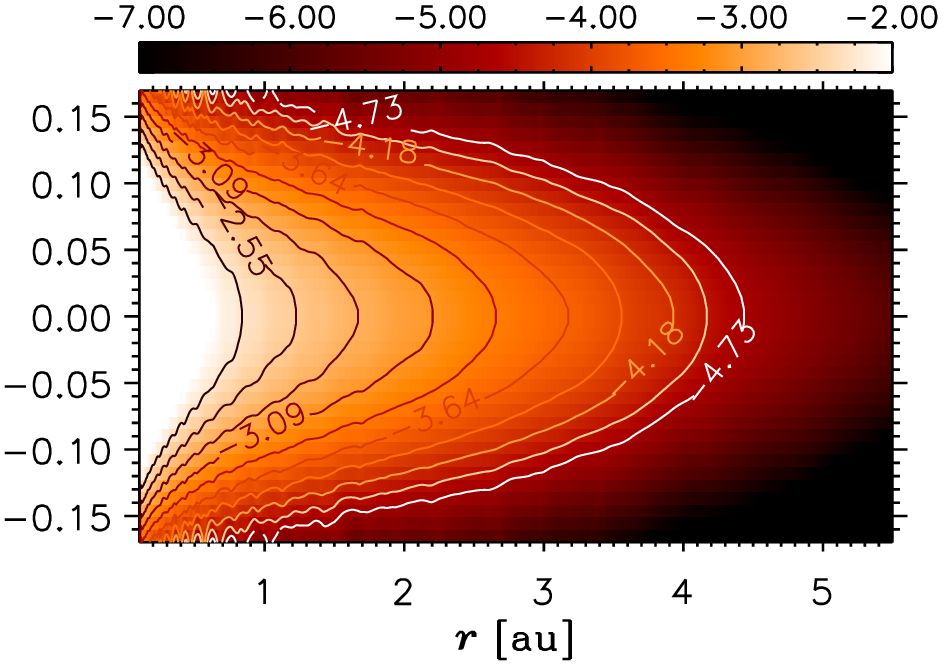}%
                          \includegraphics[clip]{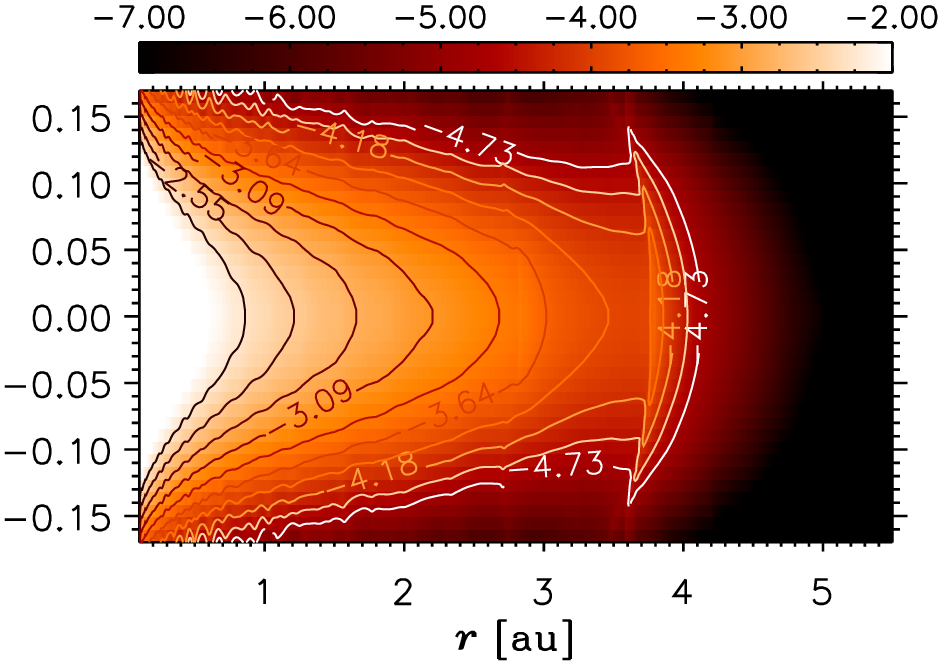}}
\resizebox{\linewidth}{!}{\includegraphics[clip]{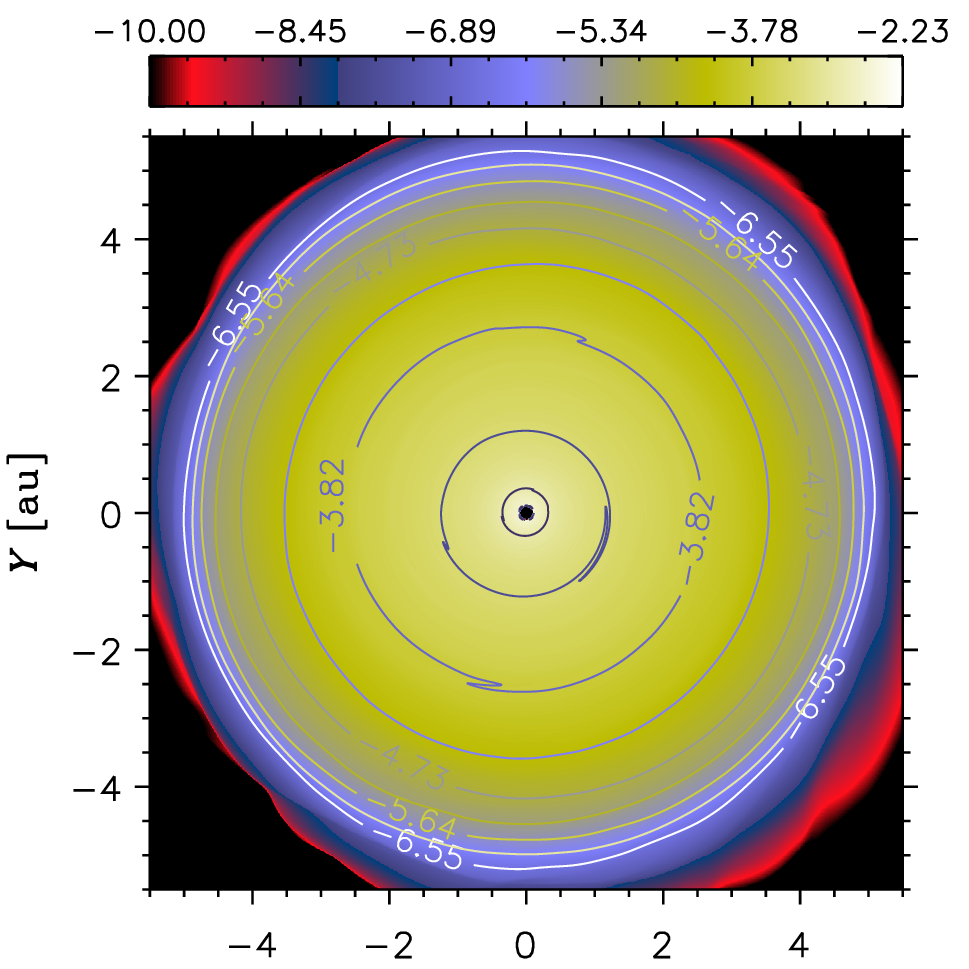}%
                          \includegraphics[clip]{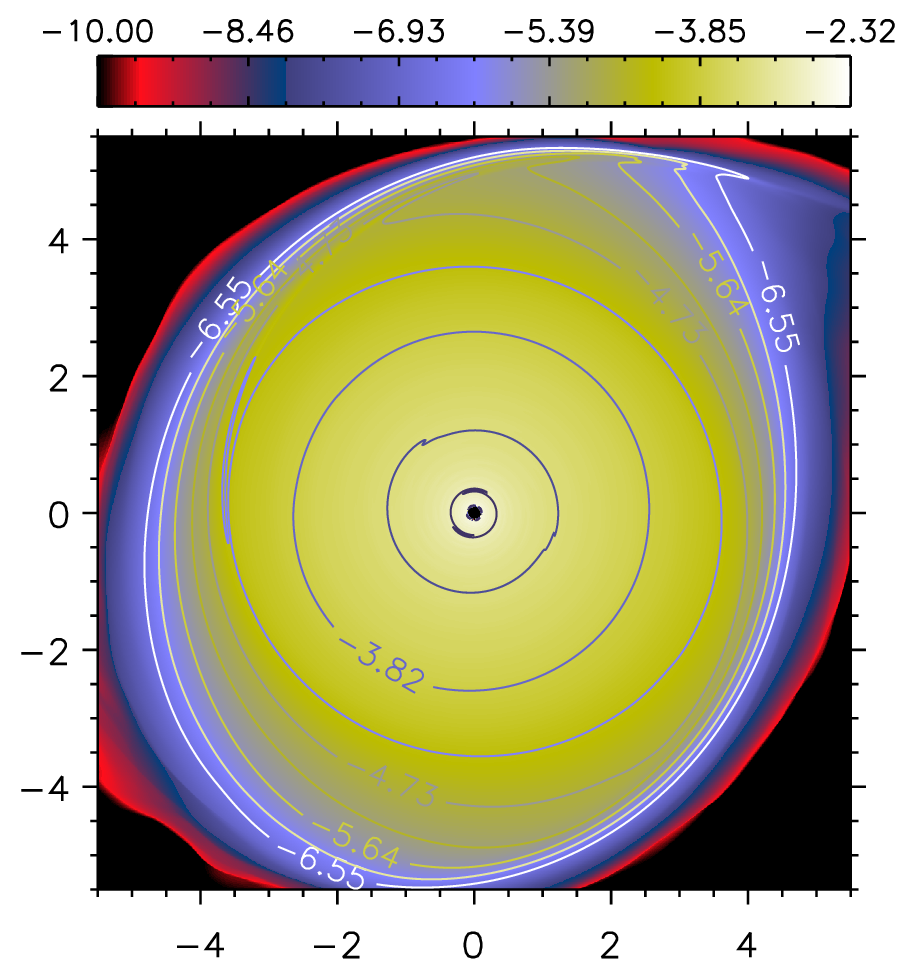}%
                          \includegraphics[clip]{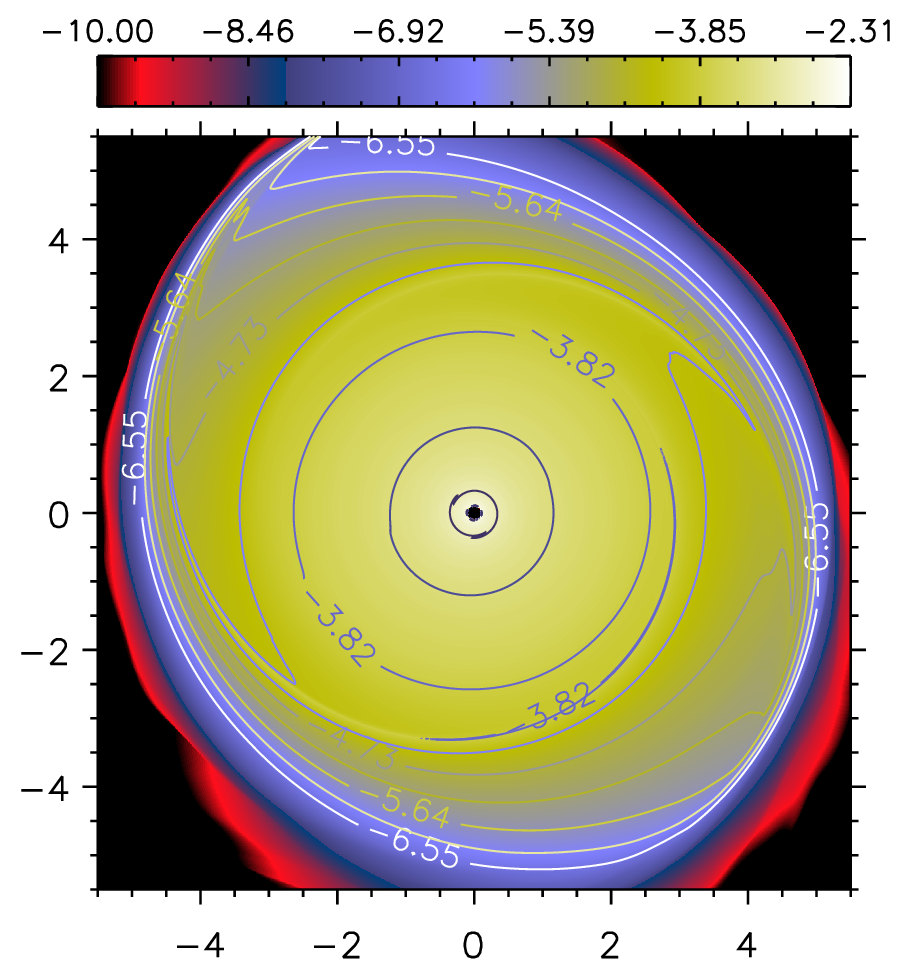}}
\resizebox{\linewidth}{!}{\includegraphics[clip]{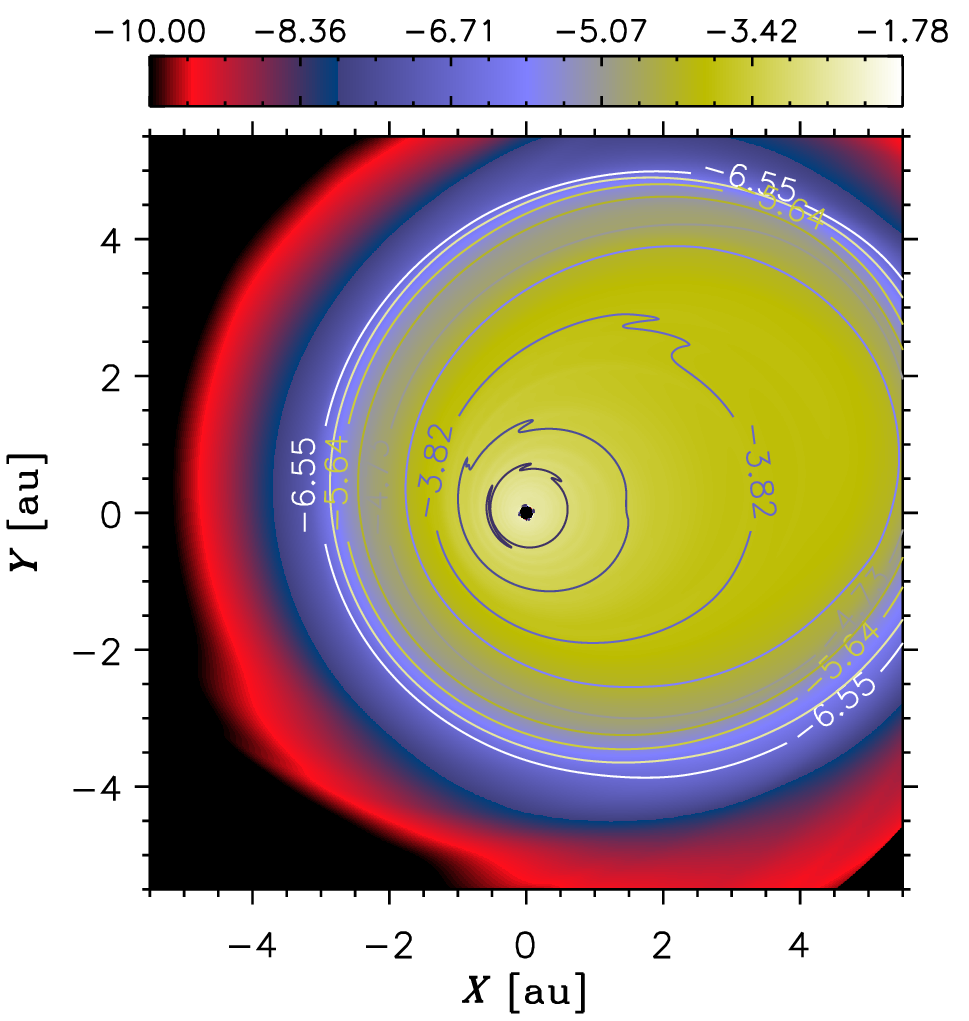}%
                          \includegraphics[clip]{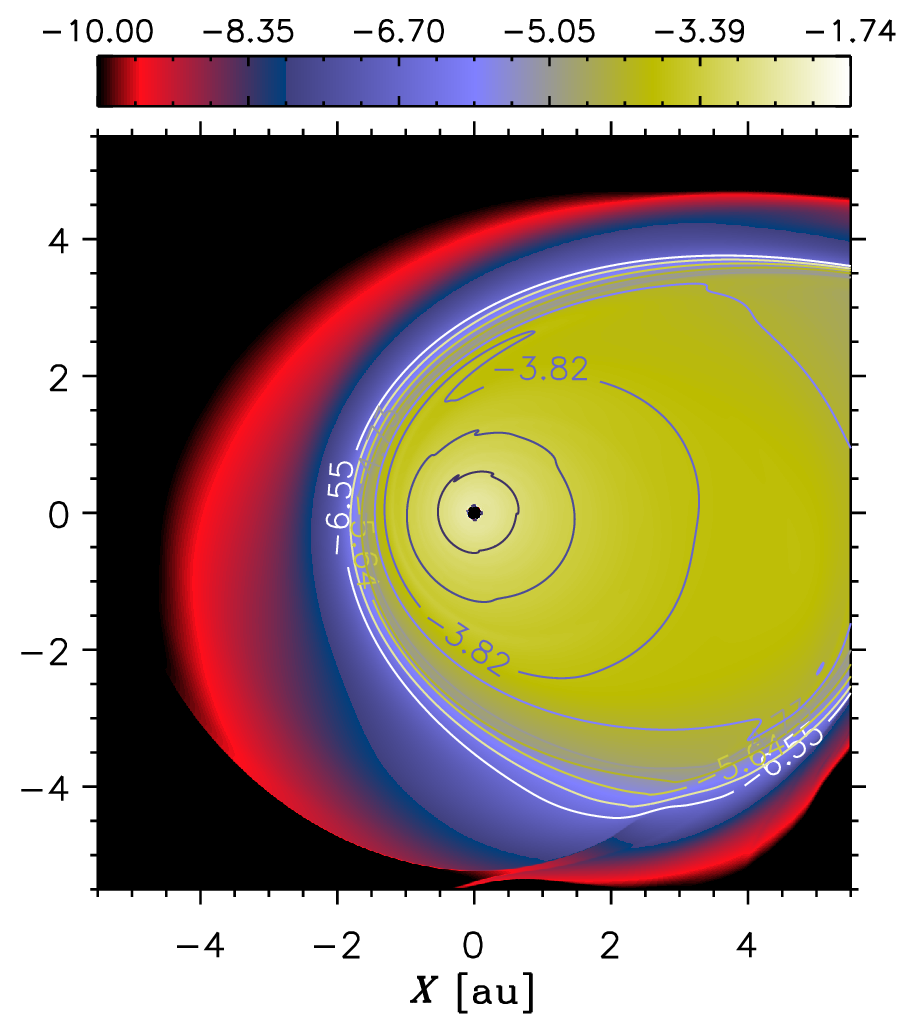}%
                          \includegraphics[clip]{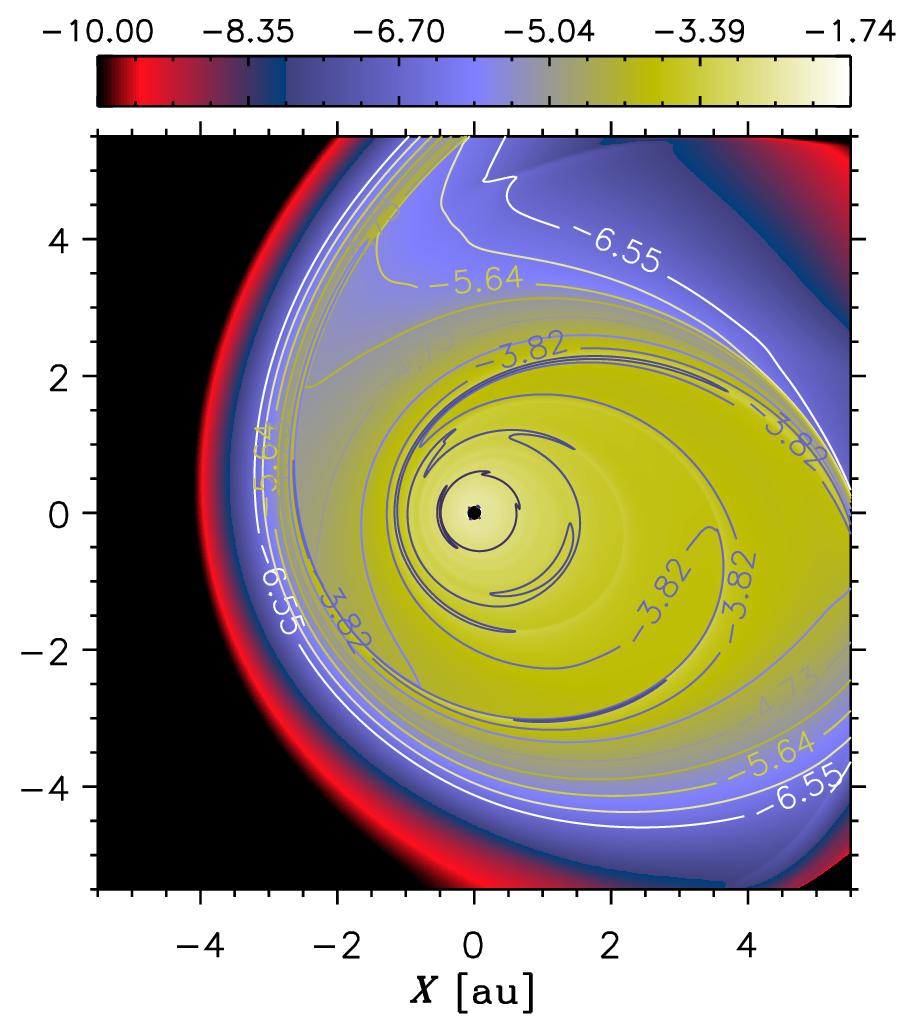}}
\caption{%
        Top panels: Volume gas density in a vertical slice ($r$-$\theta$),
        in units of $\MA/\AU^3$ and logarithmic scale, obtained
        from a 3D model. The distributions are displayed at 
        a phase of the binary orbit around apocenter passage (left),
        pericenter passage (center), and shortly thereafter (right).
        Middle and bottom panels: Surface density of the gas,
        in units of $\MA/\AU^2$ and logarithmic scale, at the same
        orbital phases as in the top panels, obtained from the 3D model
        (middle) and from a corresponding 2D model (bottom).
        }
\label{fig:3d_2d}
\end{figure*}

\begin{figure}
\centering%
\resizebox{\linewidth}{!}{\includegraphics[clip]{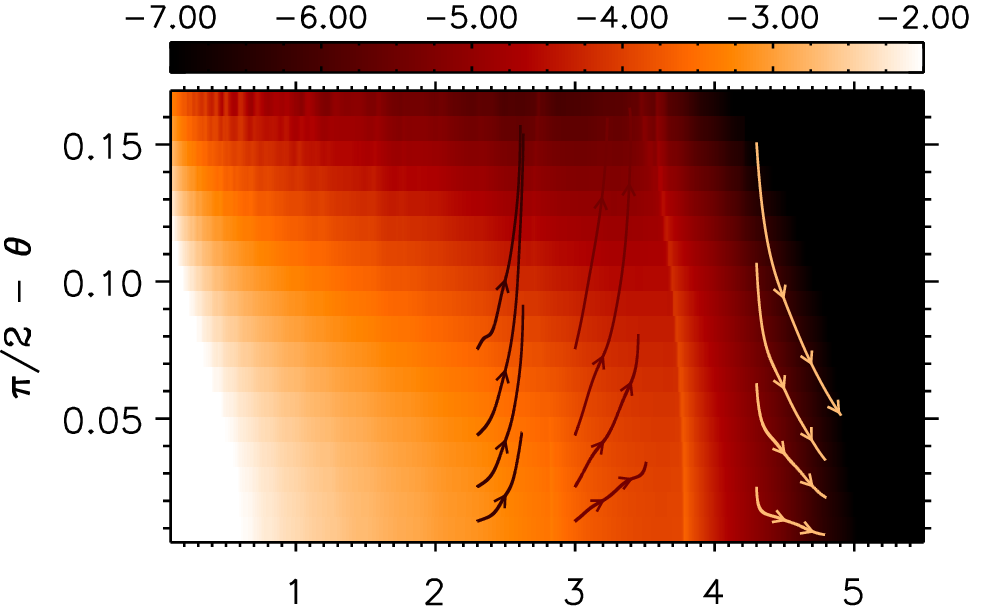}}
\resizebox{\linewidth}{!}{\includegraphics[clip]{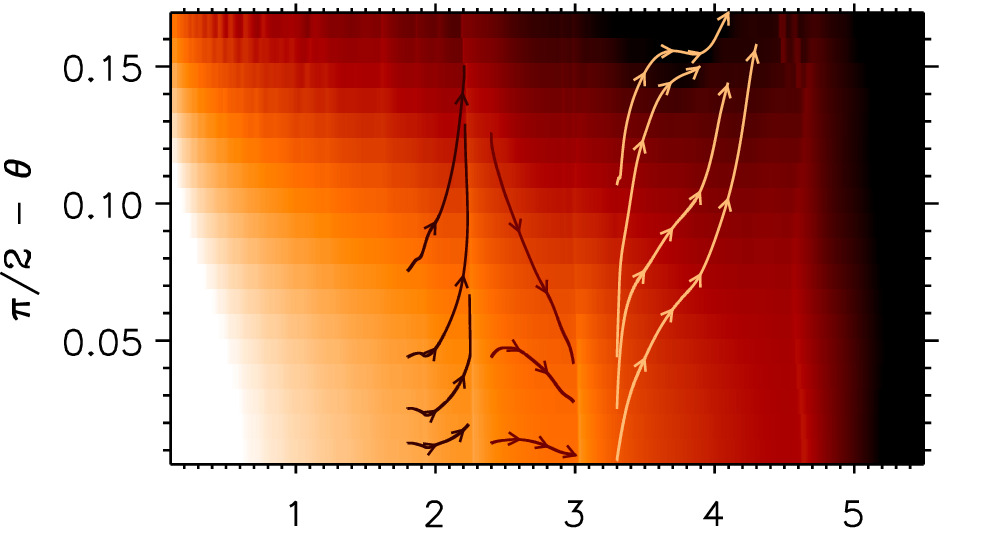}}
\resizebox{\linewidth}{!}{\includegraphics[clip]{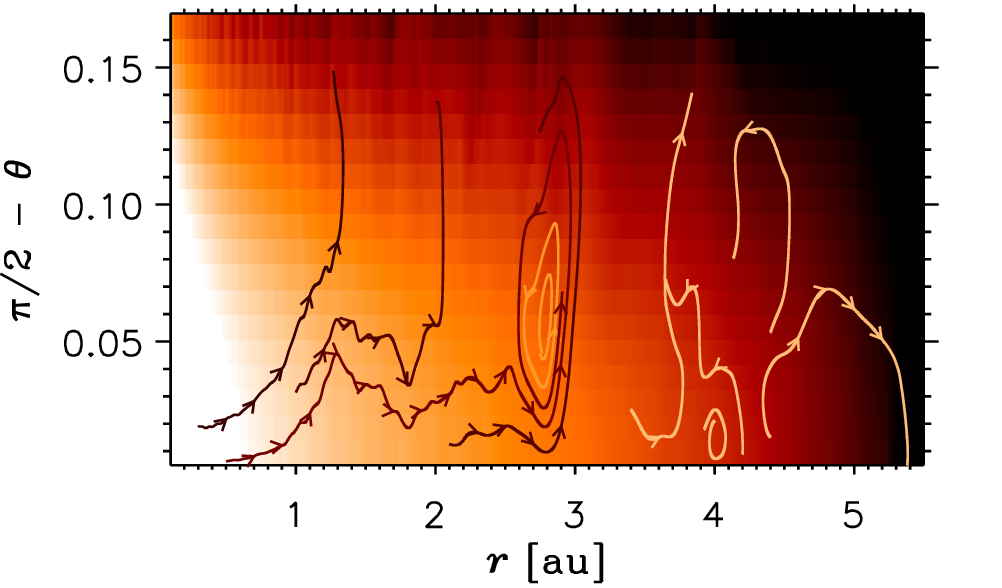}}
\caption{%
        Volume gas density in a $r$-$\theta$ plane, in units of 
        $\MA/\AU^3$ and logarithmic scale, from the 3D model of
        \cifig{fig:3d_2d}. The orbital phase of the binary corresponds
        to a time $\approx 0.06\,T$ (top) and $\approx 0.13\,T$ (middle),
        after pericenter passage, and $\approx 0.1\,T$ prior to apocenter
        passage (bottom). 
        The curves with arrows are 
        streamlines of the gas velocity projected in the $r$-$\theta$
        plane (different streamline colors are used to enhance 
        the contrast against the density color-map). 
        The waves excited by the secondary propagate inward,
        stirring the vertical and radial motion of the gas.
        }
\label{fig:3d_st}
\end{figure}

The circumprimary disk of close and eccentric binaries
is expected to be very small. The truncation radius occurs in the proximity
of the inner 1:8 mean-motion resonance of the binary, $\approx 5\,\AU$ from
the primary in the case of \gcep.
Nonetheless, outward viscous diffusion 
\citep[see][]{lynden-bell1974,pringle1981} would be inhibited 
or largely prevented by the tidal field of the stars, preserving 
a significant reservoir of material into a compact disk. 
For the initial conditions applied herein, the circumprimary disk mass
is $\approx 0.01\,\MA$.
This material would be depleted via accretion onto the primary star
and possibly replenished via accretion from a circumbinary disk
\citepalias{marzari2025}. 
The disk lifetime would be determined by these two competing processes
(photo-evaporation and surface winds may also contribute).

\cifig{fig:3d_2d} illustrates the gas density $\rho$ in the $r$-$\theta$
plane (top panels) and the surface density $\Sigma$ (middle panels), 
after $\approx 600$ binary periods ($T$), when the disk has achieved 
a quasi-equilibrium state (see discussion below). 
The color scale is logarithmic; $\rho$ is in units of $\MA/\AU^3$  and
$\Sigma$ is in units of $\MA/\AU^2$.
The three images on each row refer to different phases of the binary
orbit: apocenter passage (left), pericenter passage (center),
and soon after pericenter passage (right), when the disk is perturbed
the most.
The binary orbit is fixed in the models that do not include a giant
planet, because the gravity of disk material is neglected, and the setup
is such that the secondary transits apocenter at time $t=n\,T$, 
for any integer $n$. As a reference, when the secondary is at apocenter,
the orbital phase is $\pi$ (the star lies on the negative $X$ axis 
in the left panels of \cifig{fig:3d_2d}).
In the figure (right), the orbital phase is $\approx \pi/3$, which occurs 
$\approx 0.06\,T$ after pericenter passage (center panels).
Since the deformation of the disk depends on the phase of the binary, 
the gas density in a vertical slice depends on the azimuth around 
the primary at which it is selected. In the top panels of the figure,
the azimuth angle of the slice is $\approx 0.3\pi$ in all three cases.
Therefore, in the top-right panel, the slice is nearly aligned with 
and in the same direction as the secondary star.

The bottom panels of \cifig{fig:3d_2d} show the surface density resulting
from a corresponding 2D model, at the same times as in the middle panels.
The tidal deformation of the 2D disk is significantly larger than that of
the 3D disk, as can be visually estimated by comparing the density contours
in the middle and bottom images.
This outcome stems from the fact that, in the 2D geometry, all disk
material is confined to the orbital plane of the binary, where 
the tidal field is the strongest.

Tidal deformation and density wave propagation would also modify 
the local temperature of the gas in the affected regions. 
These variations are not modeled here because of the applied 
local-isothermal equation of state.
However, when averaged over the binary period (or longer times),
temperature asymmetries
around the primary tend to cancel out \citep[see, e.g.,][]{picogna2013}.

Density waves excited around pericenter passage propagate inward, 
from the edge of the circumprimary disk, inducing perturbations that affect
the entire vertical column of the disk, as shown in \cifig{fig:3d_st}. 
The traveling waves stir the vertical motion of the gas, ahead and behind 
of the wave fronts, as indicated by the projected streamlines 
over-plotted to the density distribution. The images represent phases 
soon after pericenter passage (top and middle) and soon prior 
apocenter passage.
The behavior of the streamlines suggests the occurrence of large-scale mixing
throughout the disk.
Vertical stirring of the gas and eddies with a vertical extent $\gtrsim H$ 
continue after pericenter passage.
It is expected that this mixing motion be reflected in the motion of 
dust grains that are well coupled to the gas. As a result, vigorous radial 
and vertical mixing of small dust grains may ensue.

\begin{figure}
\centering%
\resizebox{0.986\linewidth}{!}{\includegraphics[clip]{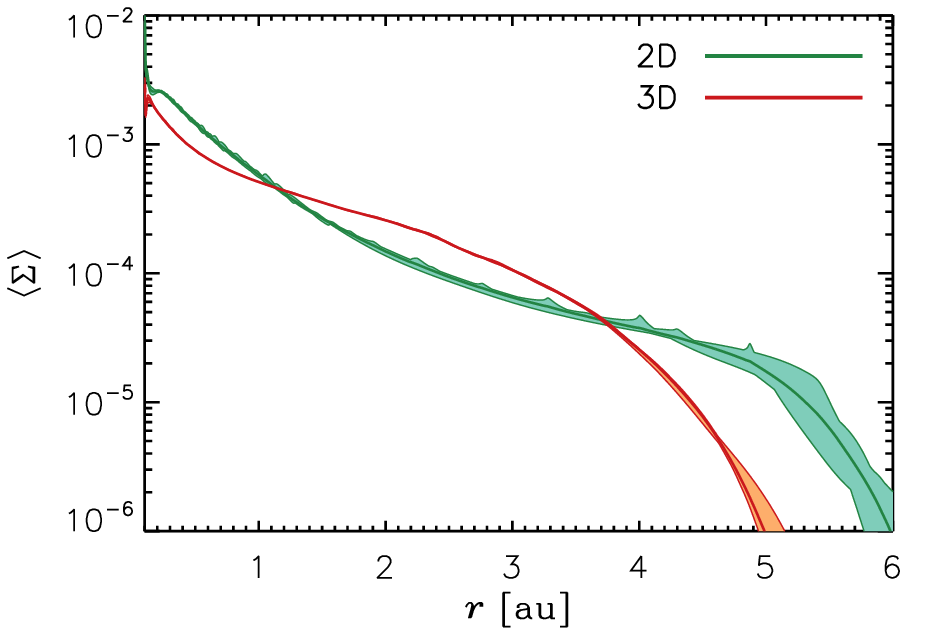}}%
\caption{%
        Surface density averaged around the primary, in units of
        $\MA/\AU^2$, obtained from the 3D (red) and 2D (green) models
        of \cifig{fig:3d_2d}. The thicker curves represent the mean
        profile computed over an orbital period of the binary, $T$.
        The shaded regions indicate maximum and minimum density values
        attained during the orbit.
        }
\label{fig:sig_av}
\end{figure}

The gas surface density averaged around the primary is reported in
\cifig{fig:sig_av}, for the 2D and 3D models of \cifig{fig:3d_2d},
as indicated. 
The thicker curves represent the mean profiles over a binary period
whereas the shaded regions provide the range of variation during that
period.
The tidal truncation radius is not well defined and, for a given
binary mass ratio and eccentricity, depends on gas pressure and 
turbulence viscosity.
Referring to the figure, this radius may be defined as the
location where $\langle\Sigma\rangle$ changes slope and declines
more rapidly.
Under this definition, truncation occurs between $\approx 0.2\,a$
(3D model) and $\approx 0.25\,a$ (2D model). 
Despite the fact that the two models in \cifig{fig:sig_av} are in very
different states, quasi-circular one and eccentric the other (as discussed below),
these predictions agree with the theoretical estimates of \citet[][]{pawel1994}.
The 2D models presented in \citetalias{marzari2025} have a shorter 
truncation radius than the 2D model in the figure, consistent with 
the fact that they applied a smaller pressure scale-height in 
the circumprimary disk \citep[see also][]{pawel1994}.

The 3D model of Figures~\ref{fig:3d_2d} and \ref{fig:sig_av} predicts that 
a negligible amount of gas is lost through the outer boundary and, therefore, 
the circumprimary disk only depletes through accretion on the primary (and 
through other possible removal processes at the surface). Viscosity-driven 
transport in a steady-state accretion disk, $3\pi\langle\Sigma\nu\rangle$ 
\citep{lynden-bell1974}, would result in an accretion rate toward the star
(averaged throughout the disk) of $\approx 10^{-7}\,\Msun\,\mathrm{yr}^{-1}$.
This analytical approximation does not, however, account for
the angular momentum transferred by the secondary during close encounters. 
In fact, the density waves, excited around pericenter passage, travel inward,
depositing angular momentum in the disk material.
Since density waves propagate inward of $r/a= 0.1$ (see \cifig{fig:3d_st}),
angular momentum deposition is expected to affect the mass transport throughout
the disk and accretion onto the primary.
Following \citetalias[][]{marzari2025}, we employ tracer particles 
to obtain fluid trajectories and determine a time-averaged mass transport
in the disk. Passive tracers are deployed in $60$ radial bins, from 
the inner boundary out to $r=6\,\AU$. Each radial bin initially contains $600$ 
particles, distributed over $2\pi$ in azimuth (around the primary) and over 
$3\,H$ above the mid-plane. 
Particle trajectories are integrated for a time $2\,T$ and their average
radial displacement during this period is used to quantify the average radial
velocity of the fluid, $\langle u_r\rangle$.
The mass flux $F=\rho \langle u_r\rangle$ (here $\rho$ is the density at 
the initial location of a tracer) is aggregated over all tracers, in each
of the initial radial bins, to obtain a time-averaged accretion rate
$\langle \dot{m}\rangle$ as a function of $r$.
Results indicate that $\langle \dot{m}\rangle$ fluctuates in the region most
affected by density waves, confirming that their perturbing effects persist
over the entire binary period. In the more quiescent inner region $r\le 1\,\AU$
($r/a\le 0.05$), $\langle \dot{m}\rangle$ has a mean value of 
$\approx -2.5\times 10^{-7}\,\Msun\,\mathrm{yr}^{-1}$, approximately twice
as large as the analytical estimate provided above for steady-state disks
(around single stars).
An additional estimate of $\langle \dot{m}\rangle$ was derived by monitoring
the time variations of the circumprimary disk mass, as gas flows across the inner
radial boundary. 
The value, $\approx -3\times 10^{-7}\,\Msun\,\mathrm{yr}^{-1}$, is consistent
with the estimate provided by the analysis of tracers.

This large value of $\langle \dot{m}\rangle$ would result in a rapid removal
of gas from around the primary, over a timescale of $\sim 10^{5}$ years
\citepalias[see also][]{marzari2025}.
Notice that, although the accretion rate depends on the choice of the initial
density, the removal timescale only depends on viscosity, pressure scale-height
and binary parameters.
Clearly, giant planet formation via core accretion would be unfeasible
over such brief timescales. A steady supply of material
from a circumbinary disk would be required for such a mechanism to operate.
A much smaller (by factors of tens) turbulence viscosity could extend
the lifetime of the disk. However, it would not alleviate the problem of
the limited reservoir of solids orbiting the primary, which may hinder
the assembly of a large enough planetary core required to form a giant
planet. In fact, this latter requirement would argue in favor of the concurrent
presence of a circumbinary disk and of some non-trivial turbulence level in 
the gaseous medium.

\begin{figure}
\centering%
\resizebox{0.986\linewidth}{!}{\includegraphics[clip]{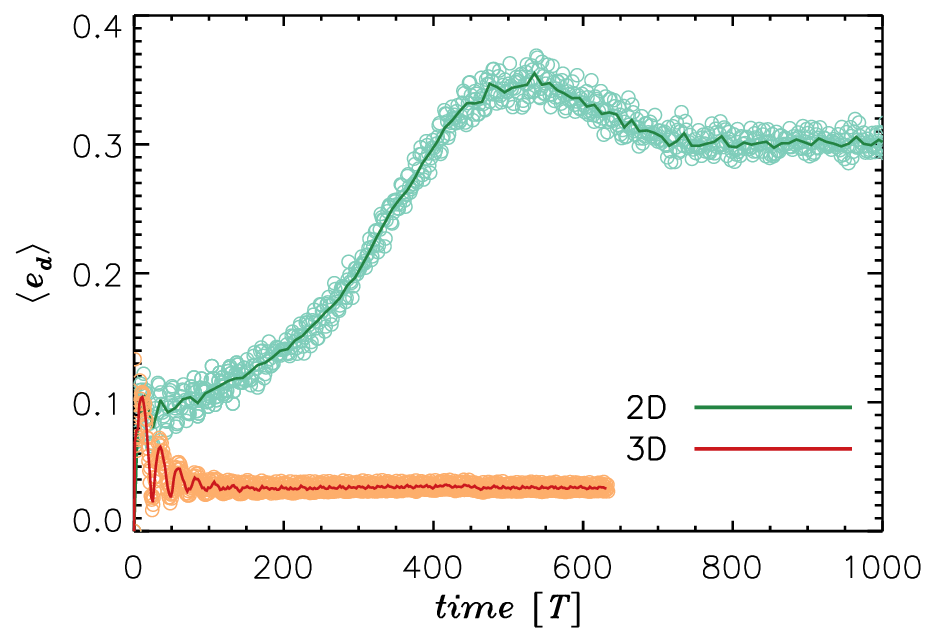}}%
\caption{%
        Mass-weighted disk eccentricity, \cieq{eq:edisk}, of
        the circumprimary disk versus time. The 2D model predicts
        an eccentric disk, achieving 
        quasi-equilibrium at $\langle e_{d}\rangle\approx 0.3$
        (green circles).
        The solid line is an average performed every $10\,T$.
        In comparison, the 3D model predicts a much less eccentric
        circumprimary disk (orange circles),
        $\langle e_{d}\rangle\approx 0.03$. The red curve represents
        an average over $10\,T$ intervals.
        }
\label{fig:ed_av}
\end{figure}

The early phases of planet formation would be aided by the largely
circular motion of the disk gas attained in the 3D models, which would tend
to reduce relative velocities among solids and promote growth. Predictions
of 2D models would instead further complicate planetary assembly.
In fact, previous 2D models indicated that, starting from a circular and
unperturbed state, the circumprimary disk settles into a quasi-equilibrium, 
eccentric state after $400$--$500\,T$ \citep{jordan2021}.
A local disk eccentricity, $e_{d}$, is often defined by projecting 
the motion of fluid elements on two-body orbits around the star.
This quantity is then mass-weighted over the disk domain to obtain
a bulk disk eccentricity
\begin{equation}
   \langle e_{d}\rangle =\frac{\int e_{d}\,dm}{\int dm},
   \label{eq:edisk}
\end{equation}
in which the mass element is $dm=\Sigma r dr d\phi$.
If the mass average is performed only over the angular coordinates,
$\langle e_{d}(r)\rangle$ provides the mass-weighted disk
eccentricity at a given radius.
In a 3D disk, the motion of a fluid element is projected on a two-body
orbit in the disk mid-plane. The results of this approximation are
validated in \cisec{sec:DDnop}, by using the trajectories of fine dust.

\cifig{fig:ed_av} illustrates the evolution of the mass-weighted disk
eccentricity, for 2D and 3D models.
Consistently with prior studies, the 2D disk settles into an eccentric
state that reaches equilibrium after several hundred binary periods.
There is a large difference between the predictions
of the 2D and 3D model, with the former projecting a quasi-equilibrium
eccentric state at $\langle e_{d}\rangle\approx 0.3$ (green circles) and
the latter resulting in a quasi-circular circumprimary disk, 
$\langle e_{d}\rangle \approx 0.03$ (orange circles).
Moreover, the 3D model attains quasi-equilibrium
much sooner, after $\approx 100\,T$.
The lines in \cifig{fig:ed_av} represent averages over $10\,T$. 
A comparison at the grid resolutions used herein (see \cisec{sec:NM})
indicates that 2D results are insensitive whereas 3D results 
at higher resolution predict a lower mass-weighted eccentricity,
$\langle e_{d}\rangle \approx 0.025$. A linear regression to model 
data during the last $\approx 100\,T$ indicates that $\langle e_{d}\rangle$
is practically constant.

As mentioned above, the quantity $\langle e_{d}(r)\rangle$ measures
the local eccentric deformation of the disk.
In the 2D model, $\langle e_{d}(r)\rangle$ rapidly increases outward,
reaching above $\approx 0.3$ at $r=1\,\AU$ and $\approx 0.5$ at
$r=3\,\AU$. In the 3D model, 
$\langle e_{d}(r)\rangle$ remains $\lesssim 0.03$ inside $4\,\AU$,
and only starts to increase around the truncation radius.
Results presented in \cisec{sec:DD}, on the orbital eccentricity
of fine dust,
can also be used to gauge the local eccentricity of the gas.

\citet{picogna2013} performed radiative SPH calculations of close
binary disks and found eccentricities around the primary between
$0.02$ and $0.06$ \citep[see also][]{martin2020}. 
These results are comparable to those of \cifig{fig:ed_av}.
There are, however, important differences between their approach and ours.
They considered a larger mass ratio, $q=0.4$, and a wider orbit, $a=30\,\AU$.
They modeled the gas thermodynamics and obtained a different aspect ratio 
of the circumprimary disk. Moreover, viscous stresses in the gas 
are not directly comparable between SPH and Eulerian hydrodynamics.

\begin{figure}
\centering%
\resizebox{0.986\linewidth}{!}{\includegraphics[clip]{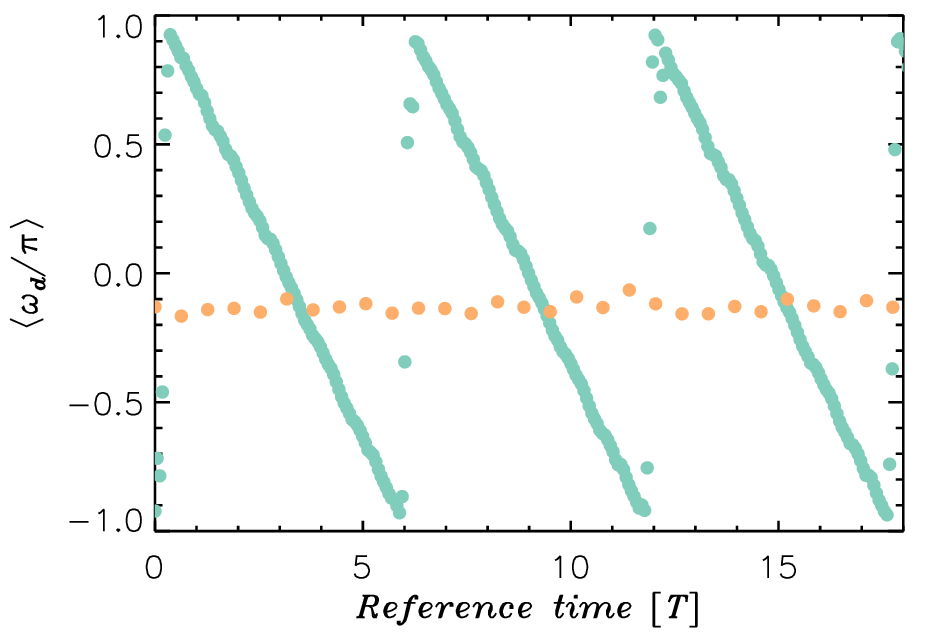}}
\resizebox{0.986\linewidth}{!}{\includegraphics[clip]{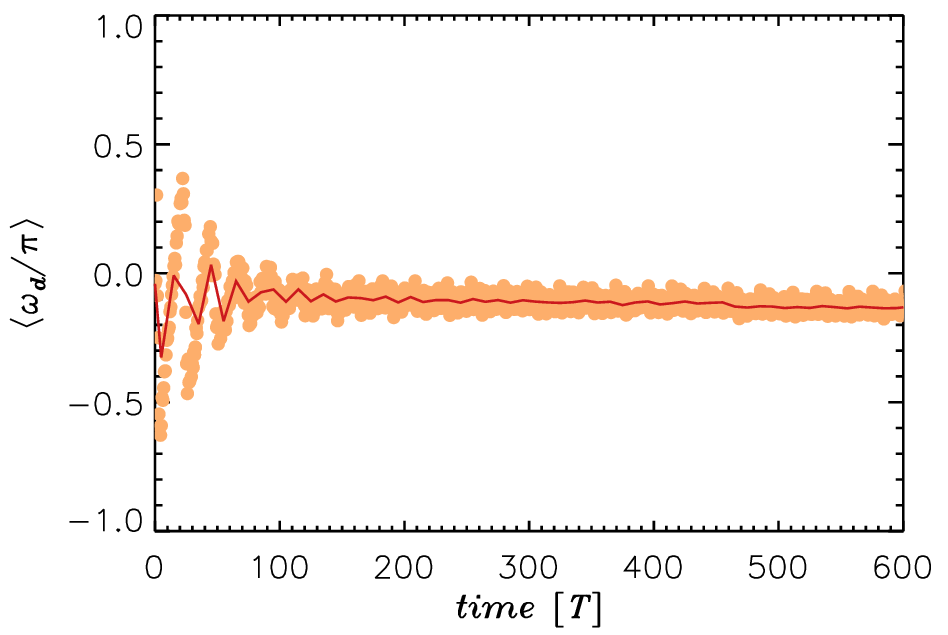}}
\caption{%
        Mass-weighted argument of pericenter of the circumprimary disk 
        versus time. The top panel shows results for a 2D (green circles)
        and a 3D model (orange circles).
        The 2D model predicts a retrograde precession of the circumprimary 
        disk at a rate of about $6$ binary periods for a full revolution.
        The precession rate resulting from the 3D model is still retrograde,
        but orders of magnitude slower.
        The long-term behavior of the 3D model is displayed in the bottom panel.
        The solid line represent an average over $10\,T$ intervals.
        See text for further details.
        }
\label{fig:apd_av}
\end{figure}

A definition similar to the mass-weighted disk eccentricity
in \cieq{eq:edisk}
can be introduced for the mass-weighted argument of pericenter 
of the disk
\begin{equation}
   \langle \omega_{d}\rangle =\frac{\int \omega_{d}\,dm}{\int dm}.
   \label{eq:apdisk}
\end{equation}
The evolution of the argument of pericenter of the circumprimary disk is
presented in \cifig{fig:apd_av} for 2D (top panel, green circles) and 3D
(orange circles) models. In the top panel, the time axis starts at some
reference time, around the end of the simulations, when the models have
already achieved equilibrium states. In 2D, the disk shows retrograde
precession at a rate $\langle \dot{\omega}_{d}\rangle\approx -\Omega/6$,
where $\Omega=2\pi/T$ is
the binary orbital frequency \citep[see also][]{jordan2021}. 
The disk precession rate is determined by a balance between 
gravitational and pressure forces, in which the former provide a prograde 
contribution to precession whereas the latter provide a retrograde contribution
\citep[e.g.,][]{lubow1992,goodchild2006}.
The 2D standard model of \citet[][]{jordan2021} has physical parameters 
comparable to those of the 2D calculation in \cifig{fig:apd_av},
although the numerical setups differ.
They also found retrograde precession, but at a lower (i.e., less negative)
rate. The difference could be related to their somewhat cooler disk 
(smaller $H/r$), which favors prograde precession.
Their results also suggest that the 2D precession rate scales as $(H/r)^{2}$,
and that it transitions from retrograde to prograde for $H/r\lesssim 0.02$ 
(for the setup they implemented, see their Fig.~17).

\cifig{fig:apd_av} also shows that, in the 3D model, the precession
rate of the circumprimary disk is much slower (top panel, orange circles) 
than the 2D estimate. The bottom panel of the figure displays the long term
evolution of the argument of pericenter in the 3D model. 
The solid line provides an average every $10\,T$. A linear 
regression for times $t>400\,T$ indicates that precession is still retrograde,
but at a rate $\langle \dot{\omega}_{d}\rangle\approx -5\times 10^{-5}\,\Omega$.
Therefore, the choice of model parameters
($\MA/\MB$, $H/r$, and $\nu$) produces a near balance between gravity and 
pressure effects. To investigate how close to prograde precession this 
configuration is, the 3D model of \cifig{fig:apd_av} was continued for 
an additional period of $\approx 250\,T$, by setting a smaller pressure 
scale-height.
To remove possible effects due to changes in kinematic viscosity, $\nu$, 
the turbulence parameter $\alpha$ was adjusted so to preserve the product
$\alpha H^{2}$.
The results of these additional models suggest that the transition from 
a retrograde to prograde circumprimary disk is very close and occurs between 
$H/r\approx 0.045$ and $0.04$. 

An eccentric gaseous disk with a significant precession rate, like the one
resulting from the 2D model discussed here, was also found by \citet{kley2008}
under similar initial conditions. Gas dynamics in such disks is expected
to have important consequences on the dynamics of planetesimals
\citep{beauge2010,silsbee2015}. In a binary configuration such as the one
we consider herein, \citet{beauge2010} showed that, on average, the relative
impact velocities between planetesimals increase with respect to the case 
of a circular, non-precessing disk. These more energetic impacts are particularly
relevant for small, kilometre-size planetesimals, and may prevent their growth
into larger bodies. Pair-wise impacts involving larger planetesimals, or 
involving two bodies with a large size difference, can still lead to accretion
but mainly in the outer regions of the disk \citep[see Fig.~7 of][]{beauge2010}.
In addition, planetesimals undergo an enhanced orbital decay in semi-major axis 
due to gas drag caused by the increase in relative velocities.
These effects would stave off, hinder or delay the formation
of a planetary core massive enough to grow into a giant planet.
However, the scenario predicted by our 3D simulations suggests that the disk
eccentricity is relatively small and the precession rate (either retrograde
or prograde) is very slow (under our assumptions).
Both occurrences would promote a more efficient
planetesimal accumulation in the circumprimary disk of close binary systems.
Other physical effects have been shown to be effective in reducing the disk
eccentricity even in 2D models, such as self-gravity \citep{marza2009}
and a better treatment of the gas thermodynamics \citep{marza2012,muller2012},
both of which are expected to be relevant in massive young disks.
\citet{jordan2021} presented 2D simulations in which they found
$e_{d} \approx 0.2$
and suggested that the low disk eccentricity found in radiative disks by
\citet{marza2012} and \citet{muller2012} may have been a consequence of low
numerical resolution and high gas viscosity. Our findings,
from the 3D simulations, indicate that a low disk eccentricity and
slow precession may be common outcomes at most stages of the circumprimary
disk evolution.

\subsection{Circumprimary disk with a giant planet}
\label{sec:GDpop}

\begin{figure*}
\centering%
\resizebox{\linewidth}{!}{\includegraphics[clip]{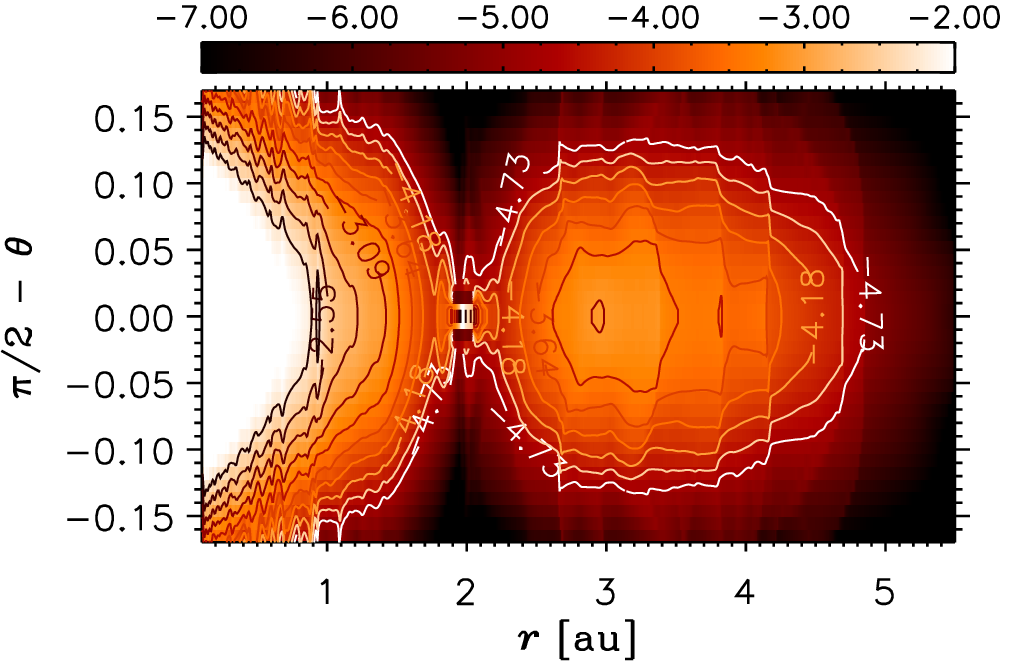}%
                          \includegraphics[clip]{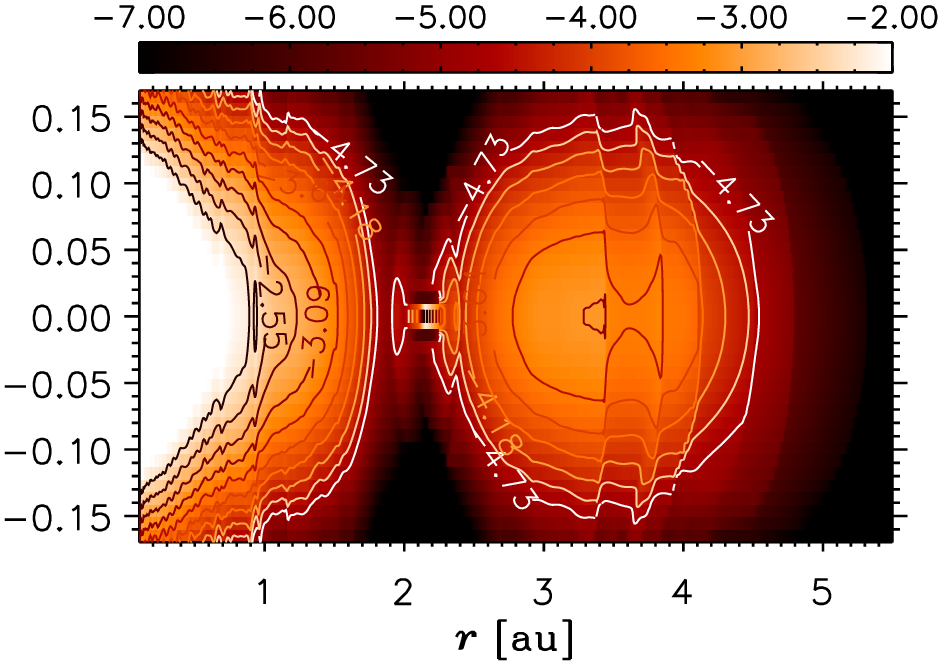}%
                          \includegraphics[clip]{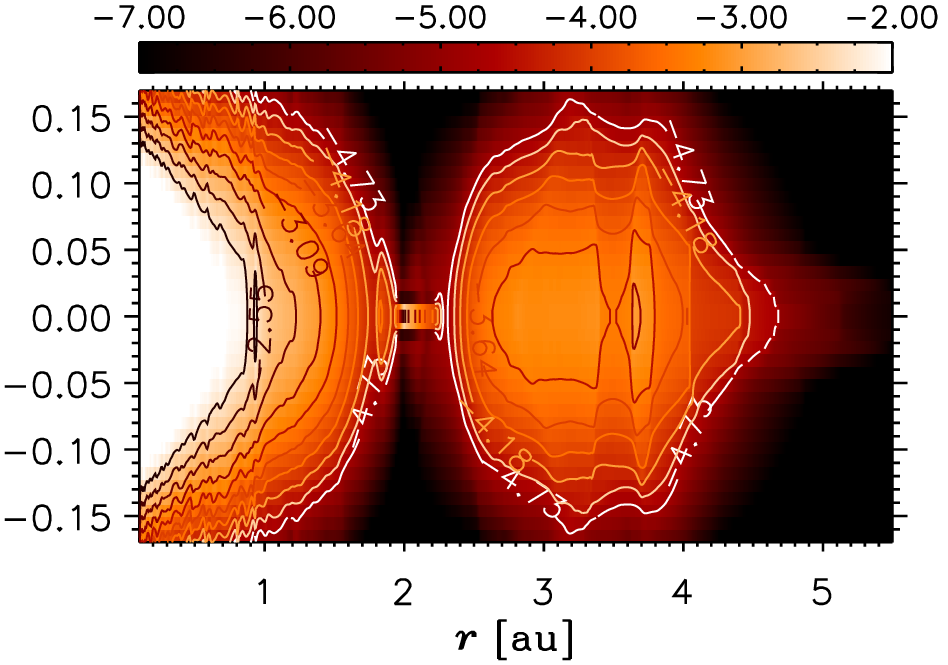}}
\resizebox{\linewidth}{!}{\includegraphics[clip]{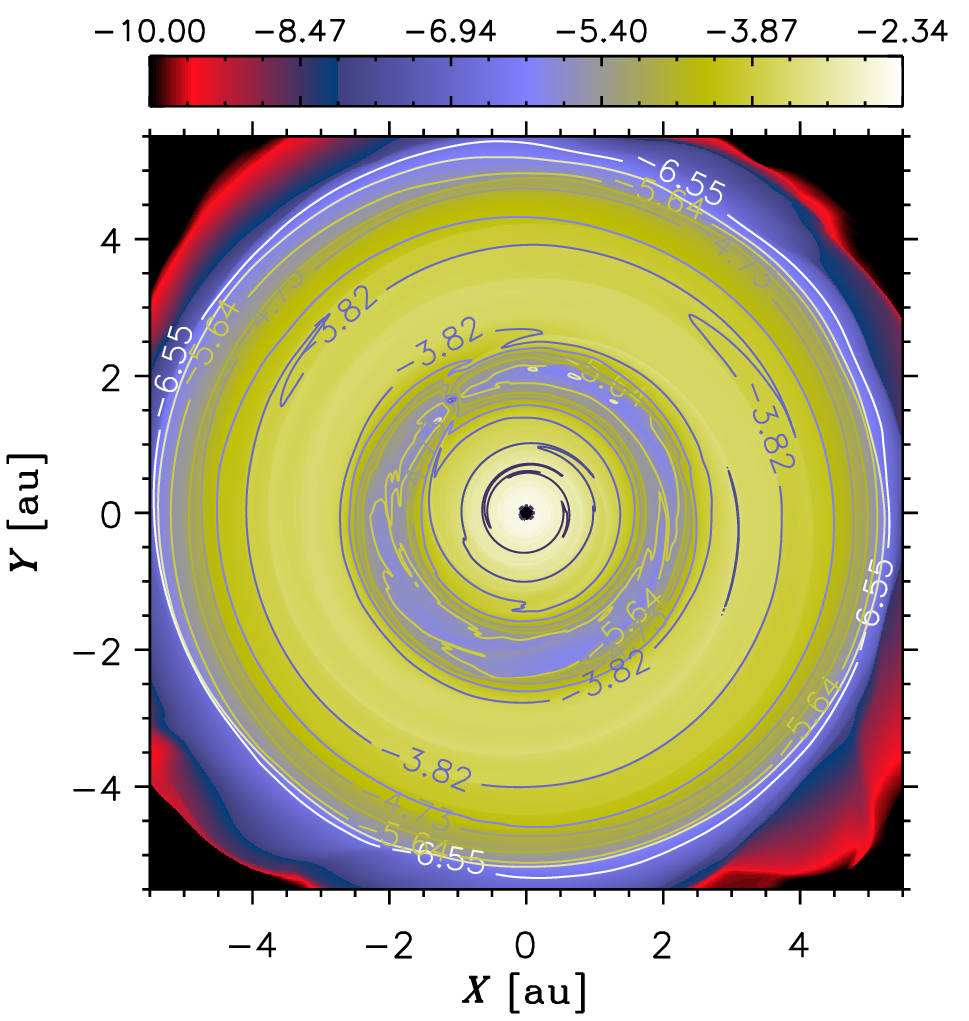}%
                          \includegraphics[clip]{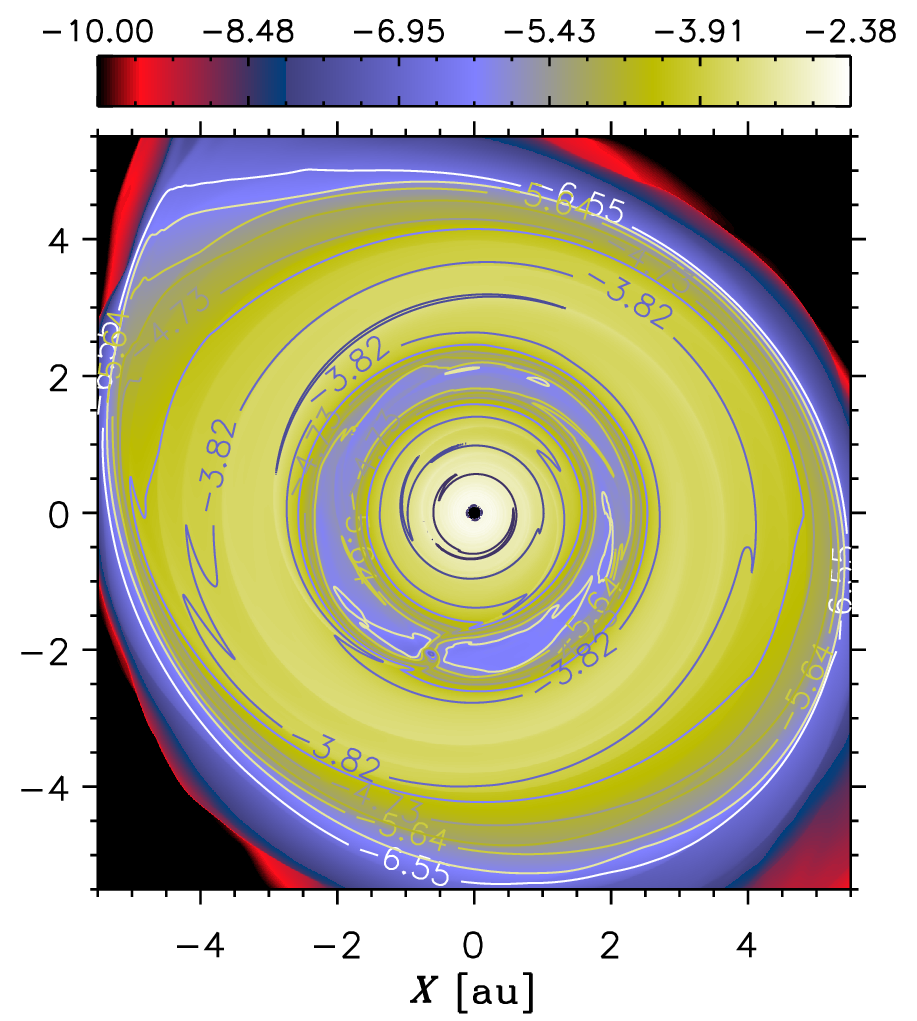}%
                          \includegraphics[clip]{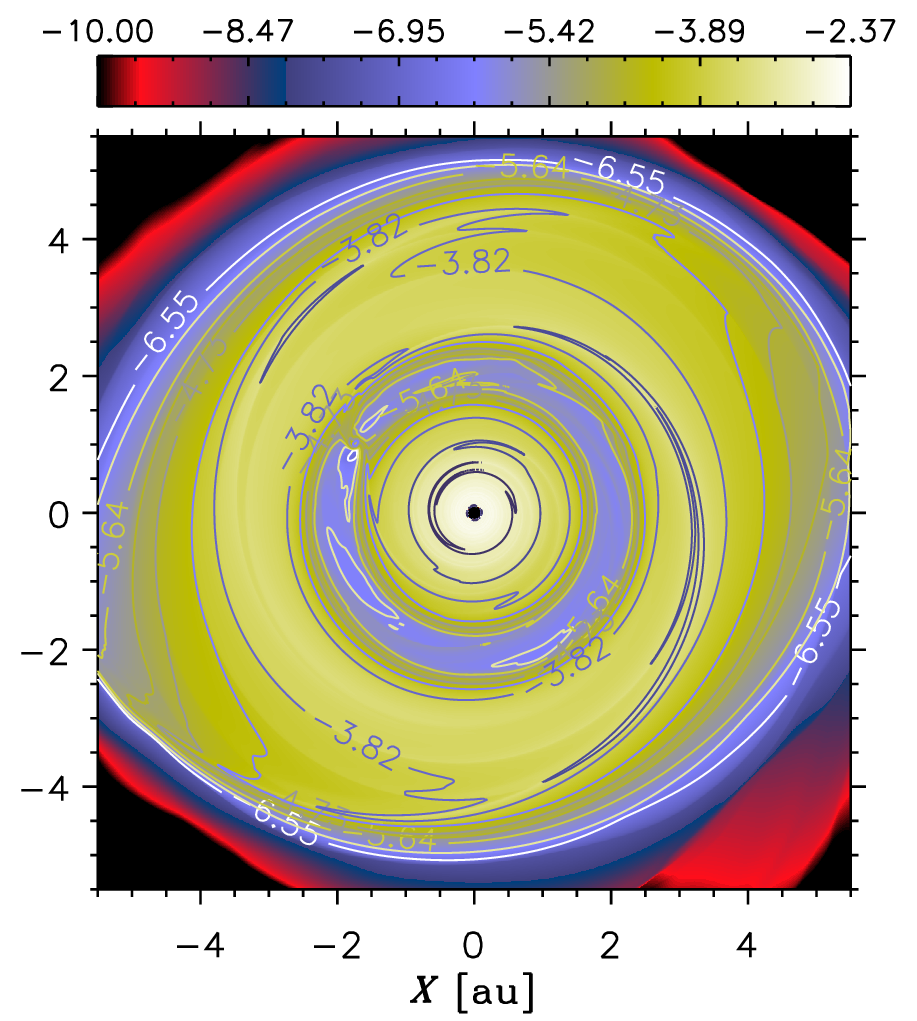}}
\caption{%
        Top panels: Volume gas density in vertical slices,
        in units of $\MA/\AU^3$ and logarithmic scale, obtained
        from a 3D model that includes a giant planet. Each
        slice passes through the planet location.
        The distributions are displayed at a binary phase around
        apocenter passage (left), pericenter passage (center),
        and shortly thereafter (right).
        Bottom panels: Surface density of the gas, in units 
        of $\MA/\AU^2$ and logarithmic scale, plotted at the same
        orbital phases as in the top panels.
        }
\label{fig:3d_wip}
\end{figure*}

\begin{figure}
\centering%
\resizebox{0.986\linewidth}{!}{\includegraphics[clip]{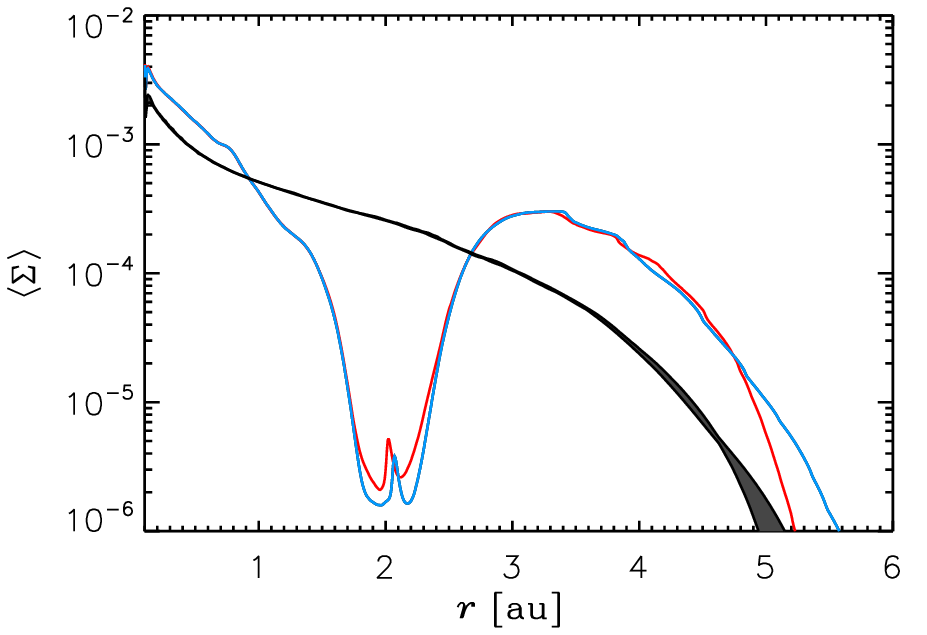}}%
\caption{%
        Surface density averaged around the primary, in units of
        $\MA/\AU^2$, obtained from the 3D model of \cifig{fig:3d_wip}.
        The density profiles correspond to orbital phases of the
        binary at apocenter (red) and pericenter (blue).
        The shaded region represents the surface density of 
        the 3D model without a planet shown in \cifig{fig:sig_av}.
        The density peaks inside the tidal gap generated by
        the planet indicate the radial position of the planet
        (which moves on an eccentric orbit).
        }
\label{fig:sig_wip}
\end{figure}

In contrast to the models presented in \cisec{sec:GDnop}, in models 
that include a giant planet orbiting the primary, the binary orbit  
varies because of stars-planet gravitational interactions. 
One consequence is that there is prograde precession of the argument
of pericenter of the binary orbit, at an average rate of
$\approx 2\times 10^{-5}\,\Omega$, where $\Omega$ is the initial orbital
frequency of the binary.
The same gravitational interactions also affect
the planet's orbit, which displays a retrograde precession of its argument
of pericenter at an average rate of $\approx -10^{-3}\,\Omega$.

Maps of the gas density in the disk with the embedded planet are
illustrated in \cifig{fig:3d_wip}. Similarly to top and middle panels of
\cifig{fig:3d_2d}, the mass density (in units of $\MA/\AU^3$ and logarithm
scale) in $r$-$\theta$ planes is shown in the top panels, in which each 
plane contains the planet, at different phases of the binary orbit:
apocenter passage (left), pericenter passage (center), and soon after 
pericenter passage (right). The bottom panels render the surface density 
(in units of $\MA/\AU^2$ and logarithmic scale) at the same orbital phases
of the binary.
Comparing the top panels of Figures~\ref{fig:3d_wip} and \ref{fig:3d_2d},
it is possible to assess the added perturbations arising from the density waves
launched by the planet to those arising from the waves launched 
by the secondary star.
Stirring of disk material induced by the planet's gravity adds to that 
induced by the secondary (see \cifig{fig:3d_st}) and, therefore, 
an increased amount of mixing in the gaseous medium should ensue.
In principle, this enhanced mixing could be reflected in the vertical
distribution of fine dust since settling could be altered. 
However, as we will discuss in \cisec{sec:DD}, in terms of global features
stirring added by the planet does not significantly impact the vertical 
distribution of dust.

The mean eccentricity of the disk, according to \cieq{eq:edisk}, has an average
value $\langle e_{d}\rangle \approx 0.03$, the same as in the configuration
without the planet. However, $\langle e_{d}\rangle$ has a somewhat larger
variability during the binary period, from $\approx 0.025$ to $\approx 0.04$.
By separating the contributions arising from the gas interior and exterior
to the planet orbit, $\langle e_{d}\rangle$ ranges between $\approx 0.015$ and 
$\approx 0.02$ inside the orbit and between $\approx 0.03$ and $\approx 0.05$
outside.

The implication is that the added planet potential is not affecting much
the eccentric perturbation on the gas, which is basically dominated by
the gravity field of the binary. However, there can be local effects, mostly 
confined to the proximity of the planet's orbit. In \cisec{sec:DD}, we also 
use the orbital eccentricity of fine dust as proxy to track the local gas
eccentricity.

As indicated in \cifig{fig:3d_wip}, the planet's potential produces a deep
gap in the disk, which is also shown in the azimuthally-averaged profiles
of surface density in \cifig{fig:sig_wip}.
A condition for gap formation \citep{gennaro2010},
\begin{equation}
\left(\frac{\Mp}{\MA}\right)^{2} > 3\pi f \alpha%
\left(\frac{H}{a_{p}}\right)^{5},
\label{eq:gap_con}
\end{equation}
is largely fulfilled by the model setup (the factor $f$ depends on 
the distribution of torques exerted by the planet on the disk).
Note that the planet-to-primary mass ratio only marginally exceeds the 
Jupiter-to-Sun mass ratio.
The depth of the gap agrees well with expectations from analytic theory and
calculations of disk-planet tidal interactions. For example, the formulations
of \citet{kanagawa2015} and \citet{duffell2015} predict a density within
the gap of $\approx 2\times 10^{-6}$, in the units of \cifig{fig:sig_wip}.

Compared to the mean density profile of the disk without a planet 
(shaded region, see \cifig{fig:sig_av}), the disk is denser outside of 
the gap region, as a result of material displaced from around 
the planet's orbit. In fact, the gap forms because the tidal field 
of the planet supplies angular momentum to material exterior to the orbit
and removes it from material interior to the orbit.
Nonetheless, as in the models of 
\cisec{sec:GDnop}, the mass flux through the outer open boundary is negligible,
and the circumprimary disk only depletes through accretion on the primary
and the planet (not considering possible removal processes at the surface).

Three-body problem dynamics places a stability limit of $\approx 3.8\,\AU$
to the planet distance from the primary \citep{holman1999}.
However, interactions with the disk material can affect this limit, and
orbital dynamics in general.
In fact, variations of the planet's orbital elements, due to the gravity
field of the stars alone, occur on extremely long timescales, 
$\sim 10^{8}\,T$ for $a_{p}$ whereas $e_{p}$ executes oscillations
in the range $0.044$--$0.064$ over $\approx 90\,T$.
Experiments conducted by allowing the planet's orbit to change under torques
exerted by the disk gas indicate that the orbit tends to shrink 
and circularize. Orbital migration occurs at an average rate 
$\langle\dot{a}_{p}\rangle\approx -0.01\,\AU/T$ whereas orbital eccentricity
damps at a rate $\langle\dot{e}_{p}\rangle\sim -0.0001/T$
(undergoing oscillations over several binary periods).
These experiments neglect tidal torques exerted by the dust on the planet,
since the models include a limited number of individual particles.

A rapid evolution of the orbital elements of the planet is expected since
the circumprimary disk extends only for about $2.5$ times the orbital radius
of the planet and is ten times as massive.
In particular, the migration timescale is short but still within the canonical
Type~II regime expected of gap-opening planets 
\citep[see, e.g.,][and references therein]{baruteau2014}.
In fact, \citet{gennaro2010} showed that orbital migration of gap-opening
planets can be described analogously to Type~I migration (of non-gap-opening
planets), through a torque density distribution that peaks at 
$\approx R_{\mathrm{H}}$ on either side of $r=a_{p}$ and is non-zero
within $\approx 3\,R_{\mathrm{H}}$ of the orbit.
Therefore, migration can be described as Type~I, but driven by the average
density within $2$-$3\,R_{\mathrm{H}}$ of the planet's orbit 
\citep[see also][]{kanagawa2018}. In \cifig{fig:sig_wip}, this average density
is about two orders of magnitude lower than the unperturbed density
(black curve).
Clearly, the model configuration would allow for inward migration of
the planet on a relatively short timescale so that the orbital parameters
in Table~\ref{table:sum} would not be final values 
(i.e., after gas disperses). 

The model configuration would also allow for continued growth
of the planet. Estimates of the accretion rate, at the viscosity level
adopted in the model and $\langle\Sigma\rangle$ in \cifig{fig:sig_wip},
indicate a value of a few times $10^{-5}\,\Mjup\,\mathrm{yr}^{-1}$
\citep[][]{lissauer2009,bodenheimer2013}. But this rate would rapidly
decline as $\Mp$ increases. A simple estimate of the planet mass
based on these rates would result in a final value of
$\approx 9\,\Mjup$, assuming a constant density $\langle\Sigma\rangle$
a few times as low as in \cifig{fig:sig_wip} to account for depletion.
Incidentally, some observations suggest that the planet is actually very massive,
over $9\,\Mjup$ \citep{benedict2018}. 
Regardless, our model setup would
be conducive to a massive giant planet, provided an external source of gas.

\begin{figure}
\centering%
\resizebox{0.986\linewidth}{!}{\includegraphics[clip]{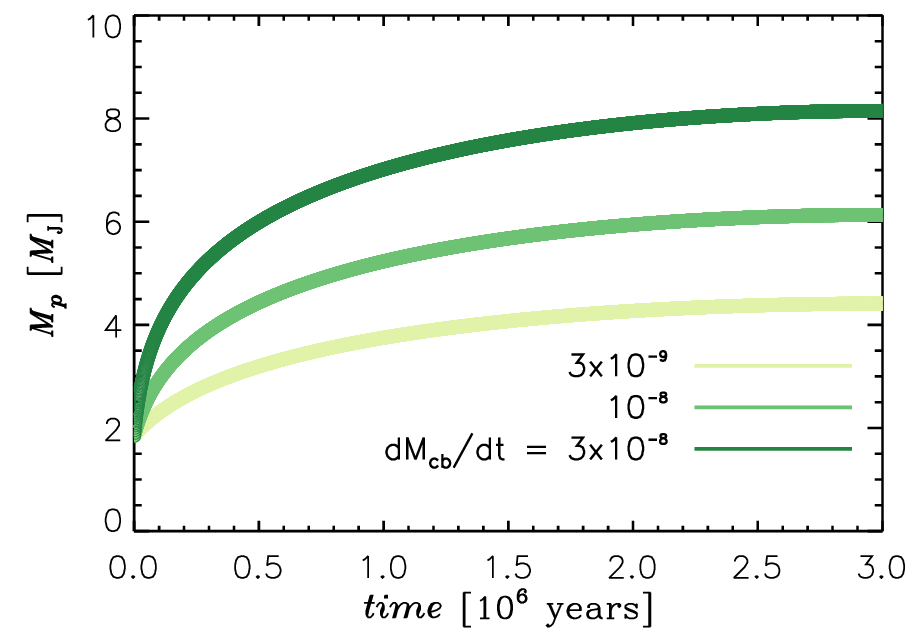}}%
\caption{%
        Integration of the planet mass according to the accretion rates
        reported in \citet[][]{lissauer2009} and \citet[][]{bodenheimer2013},
        assuming that the circumprimary disk is supplied by the circumbinary disk
        with an initial accretion rate $dM_{\mathrm{cb}}/dt$ as indicated, in units
        of $\Msun\,\mathrm{yr}^{-1}$ \citepalias[see][]{marzari2025}.
        The supply rate declines to zero over the circumbinary disk lifetime,
        a few million years in this example. See text for further details.
        }
\label{fig:fima}
\end{figure}
Mass growth would also alleviate the migration issue since the depth of the
gap region reduces as $\Mp$ increases. The Type~I migration speed is proportional 
to the planet mass, but the average density driving migration (inside 
$2$-$3\,R_{\mathrm{H}}$ of the planet's orbit) is 
$\propto \Mp^{-2}$ \citep[e.g.,][]{kanagawa2015,duffell2015},
so that the migration rate reduces with planet mass \citep[e.g.,][]{ida2018}.
However, both growth and migration of a planetary body orbiting the primary
require considerations about the lifetime of the gaseous disk. As discussed
in \cisec{sec:GDnop}, an isolated disk would be removed on a short, 
$\sim 10^{5}\,\mathrm{yr}$ timescale. A circumbinary disk would need to be
in place to supply material and prolong the disk lifetime. Over the long
term (i.e., the lifetime of the circumbinary disk), such a supply rate would
also determine the surface density in the circumprimary disk in which
the planet evolves.
According to the estimates of \citetalias[][]{marzari2025}, the circumbinary 
supply rates, $\dot{M}_{\mathrm{cb}}$, could be of order 
$10^{-8}\,\Msun\,\mathrm{yr}^{-1}$, which would imply values of 
$\langle\Sigma\rangle\sim \dot{M}_{\mathrm{cb}}/(3\pi\nu)$
several times as low as in \cifig{fig:sig_wip}.
Applying a more detailed time integration of the planet mass using functions
$\dot{M}_{p}=\dot{M}_{p}(\Mp)$ reported in 
\citet[][see the Appendix therein]{bodenheimer2013} and density values 
determined by a declining supply rate $\dot{M}_{\mathrm{cb}}$, of initial
value $\sim 10^{-8}\,\Msun\,\mathrm{yr}^{-1}$, the final mass of the planet
could reach up to $\approx 8\,\Mjup$, as indicated by the evolution
tracks reported in \cifig{fig:fima}.

\section{Dust dynamics}
\label{sec:DD}

\begin{figure*}
\centering%
\resizebox{\linewidth}{!}{\includegraphics[clip]{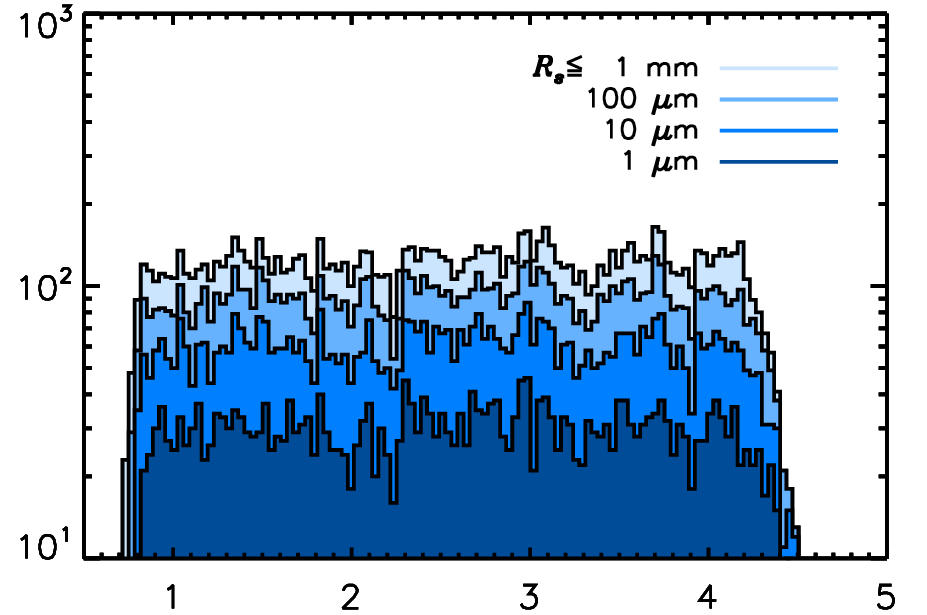}%
                             \includegraphics[clip]{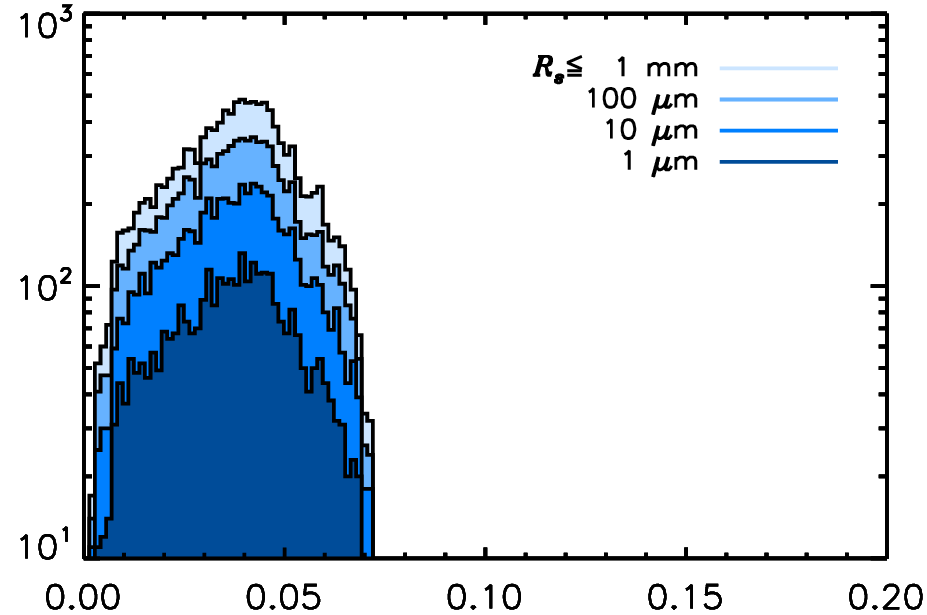}%
                             \includegraphics[clip]{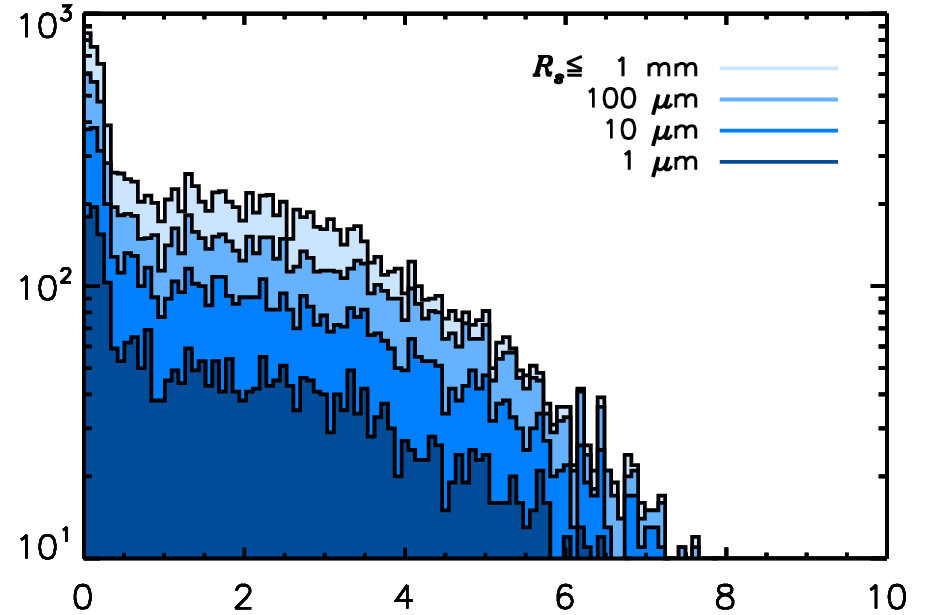}}
\resizebox{\linewidth}{!}{\includegraphics[clip]{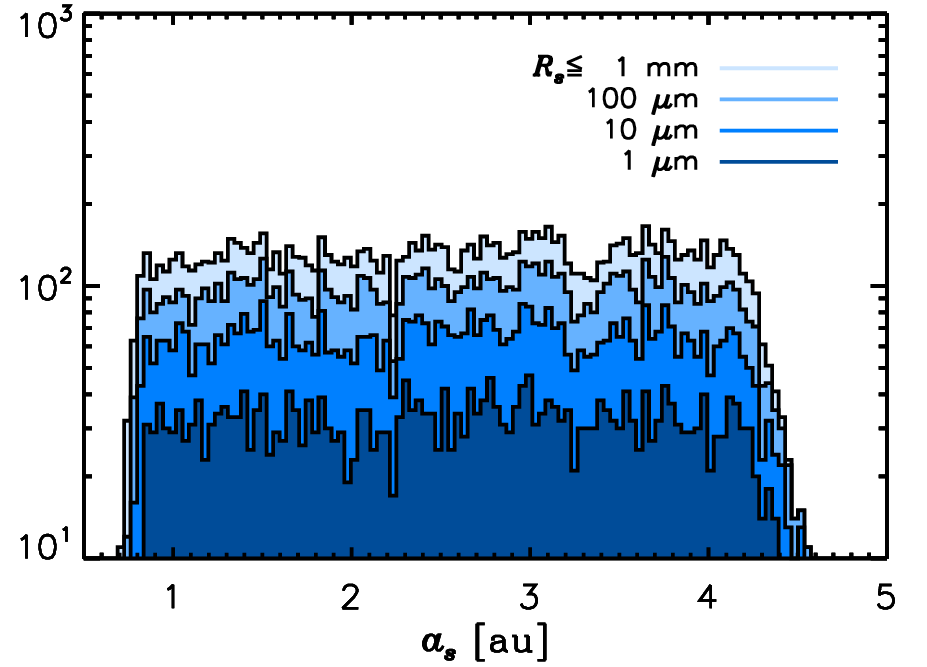}%
                             \includegraphics[clip]{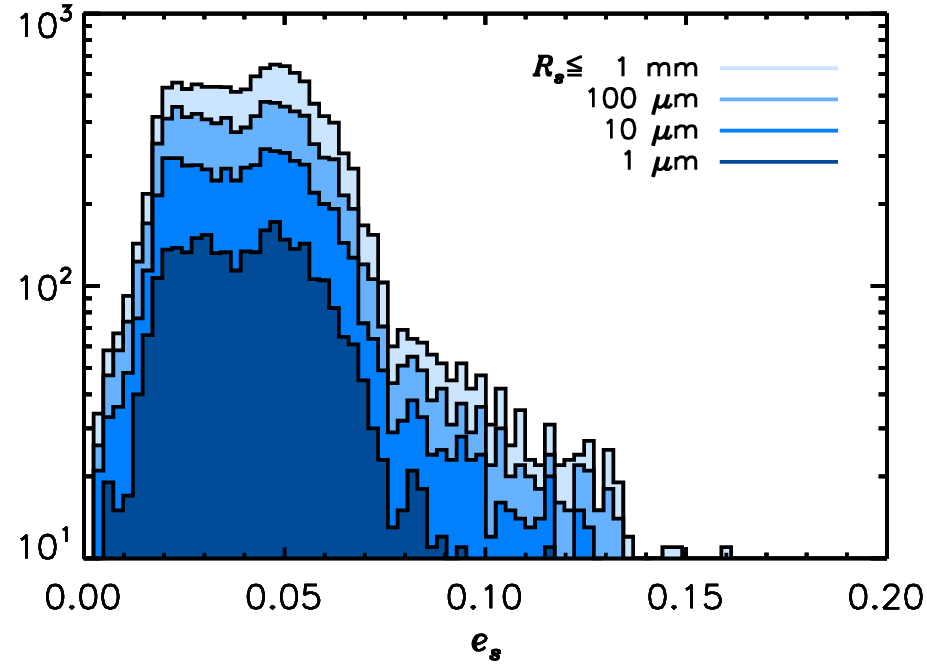}%
                             \includegraphics[clip]{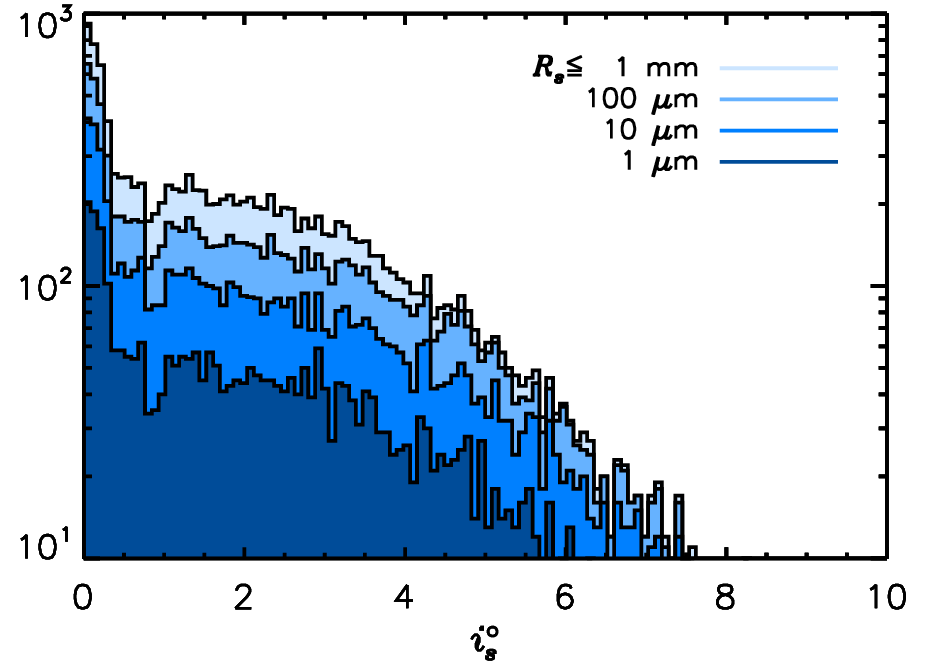}}
\caption{%
         Histograms of dust orbital properties obtained from a 3D model
         without a giant planet. Semi-major axis (left), eccentricity 
         (center), inclination (right) of the solids are displayed
         at a binary phase around apocenter passage (top) and pericenter 
         passage (bottom).
         Histograms are stacked in order of increasing dust size,
         as indicated in the legends.
        }
\label{fig:hist_nop}
\end{figure*}

\begin{figure}
\centering%
\resizebox{\linewidth}{!}{\includegraphics[clip]{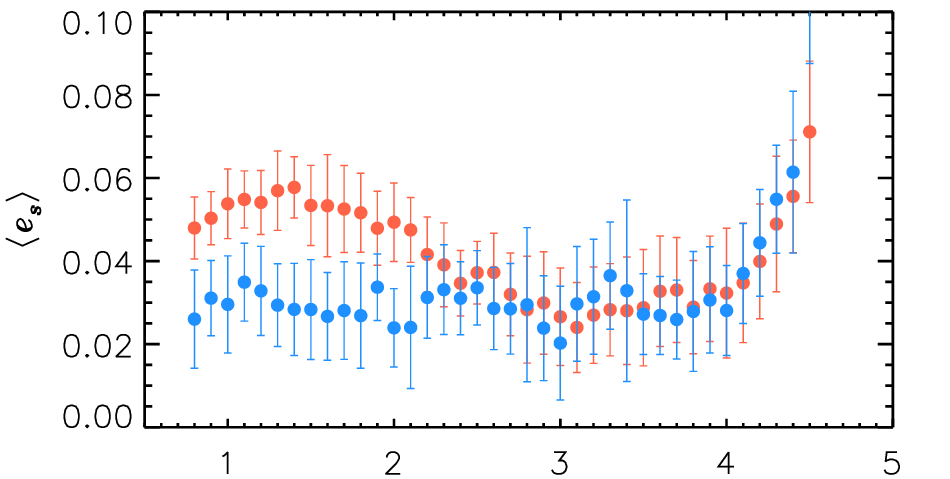}}
\resizebox{\linewidth}{!}{\includegraphics[clip]{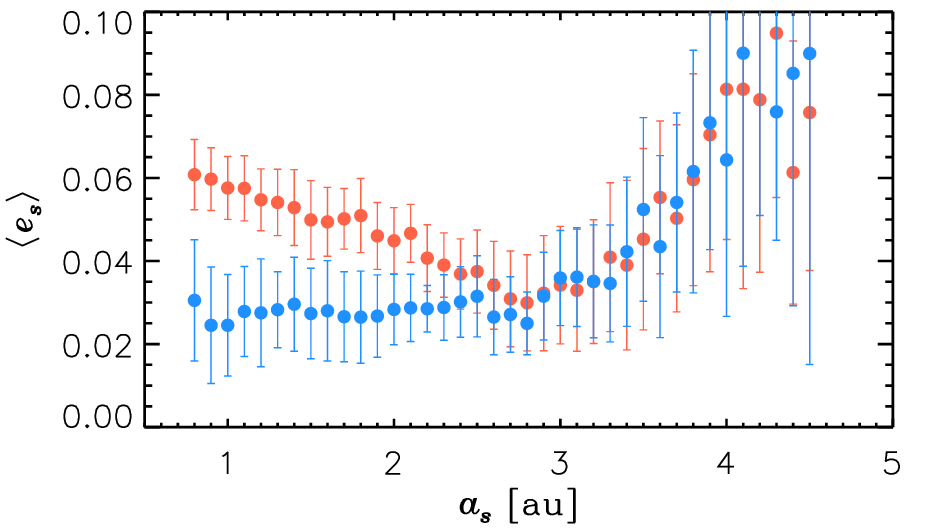}}
\caption{%
        Average orbital eccentricity of $1\,\mu\mathrm{m}$-size
        dust, as function of semi-major axis, corresponding to 
        the orbital
        phases of the binary of apocenter (top) and pericenter
        passage (bottom).
        Red symbols refer to grains within one scale-height
        $H$ of the disk mid-plane whereas blue symbols refer to grains
        moving above $H$. Error bars provide the spread 
        (standard deviation) of the distributions around the mean values.
        }
\label{fig:es_as_nop}
\end{figure}

\begin{figure*}[ht!]
\centering%
\resizebox{0.986\linewidth}{!}{\includegraphics[clip]{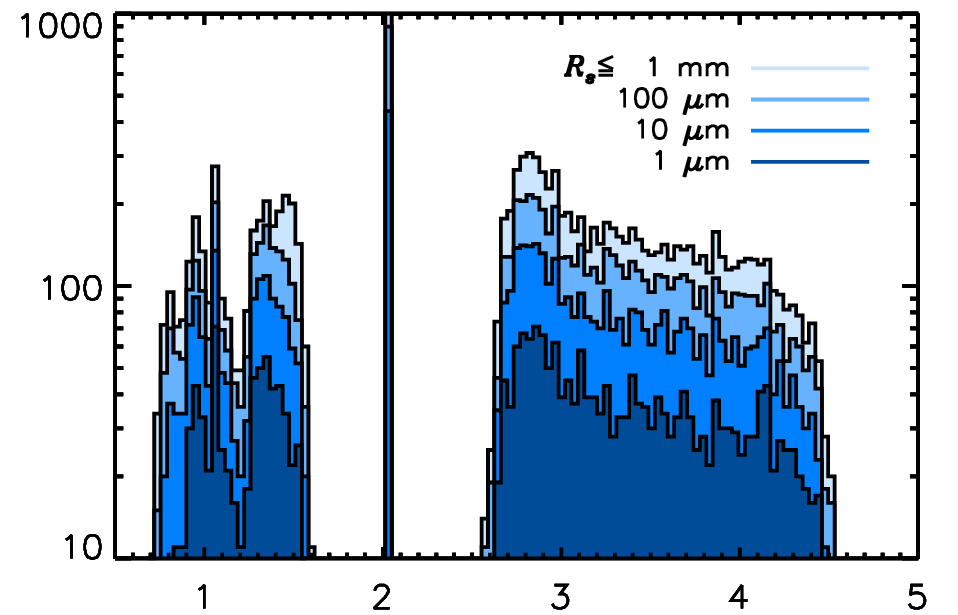}%
                             \includegraphics[clip]{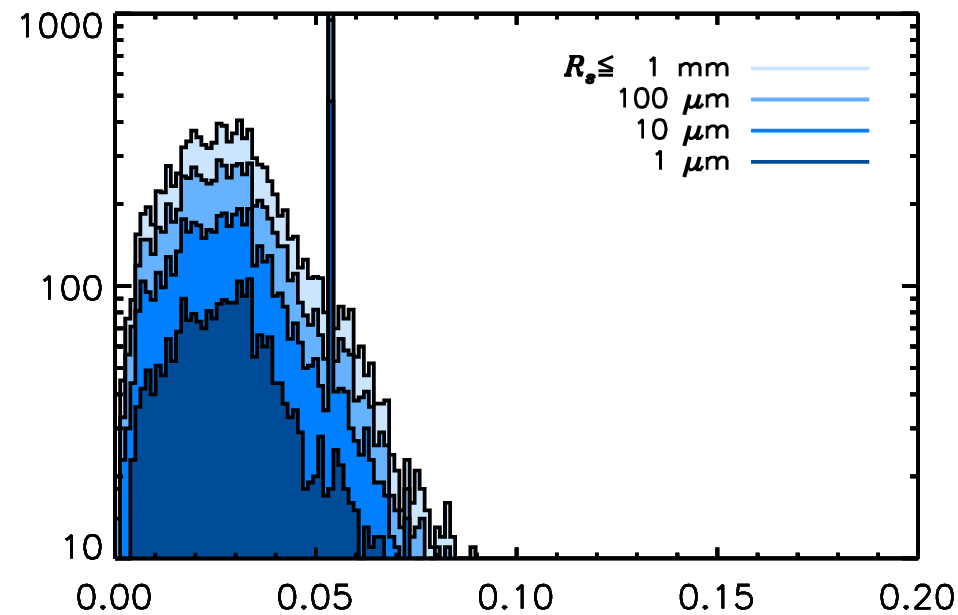}%
                             \includegraphics[clip]{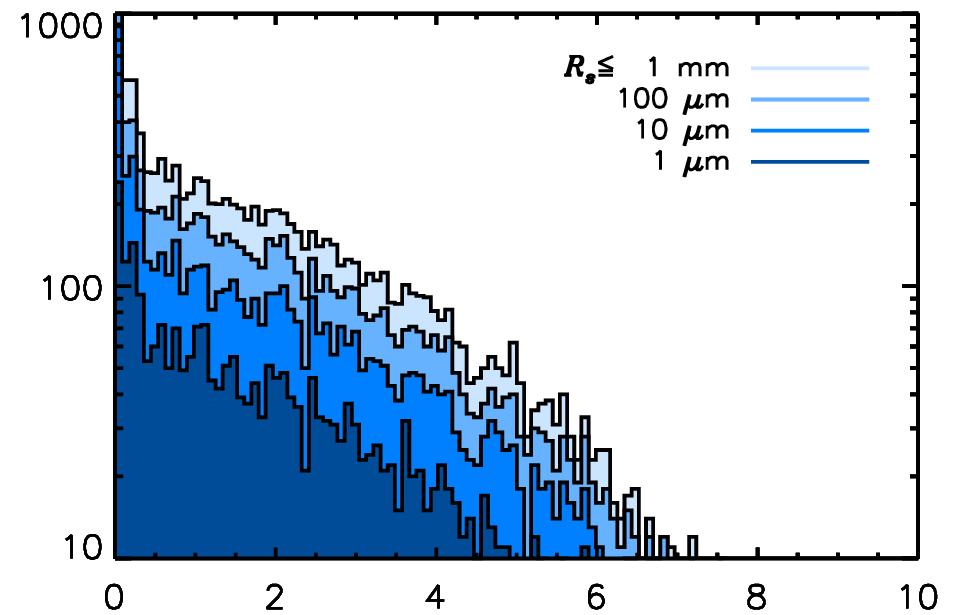}}
\resizebox{0.986\linewidth}{!}{\includegraphics[clip]{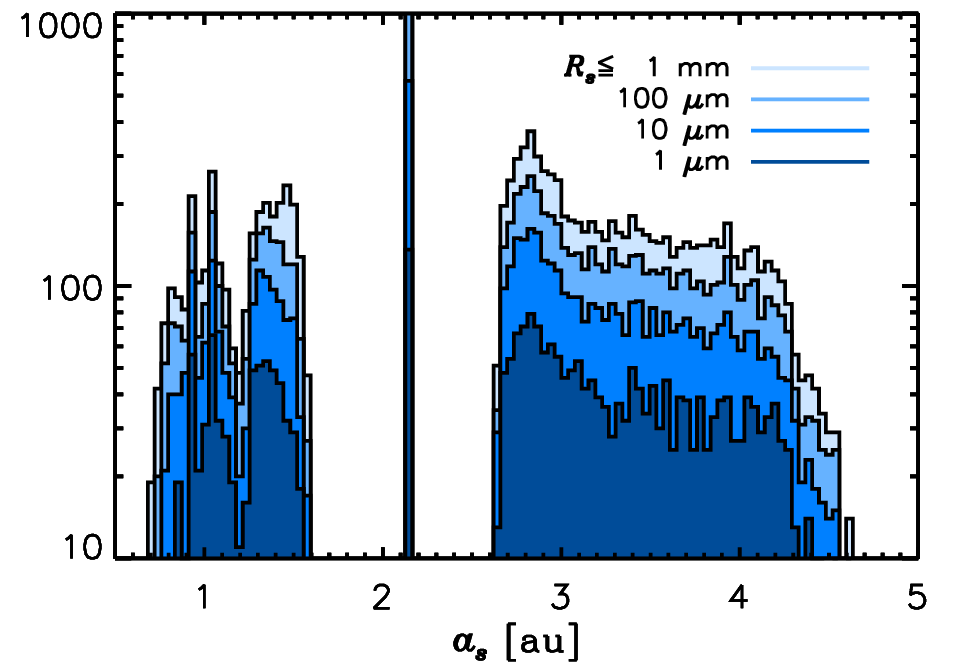}%
                             \includegraphics[clip]{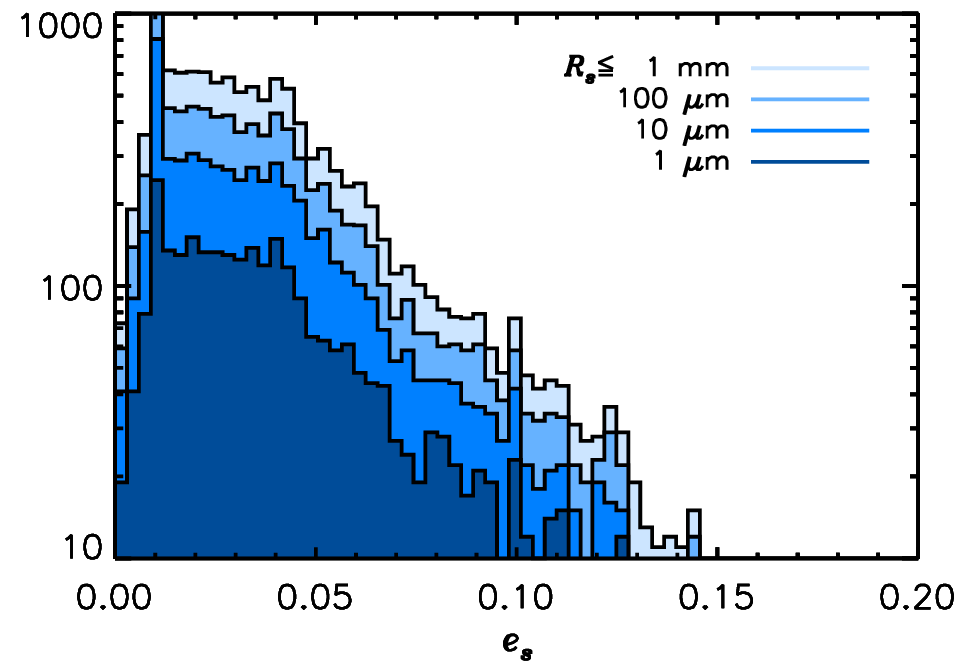}%
                             \includegraphics[clip]{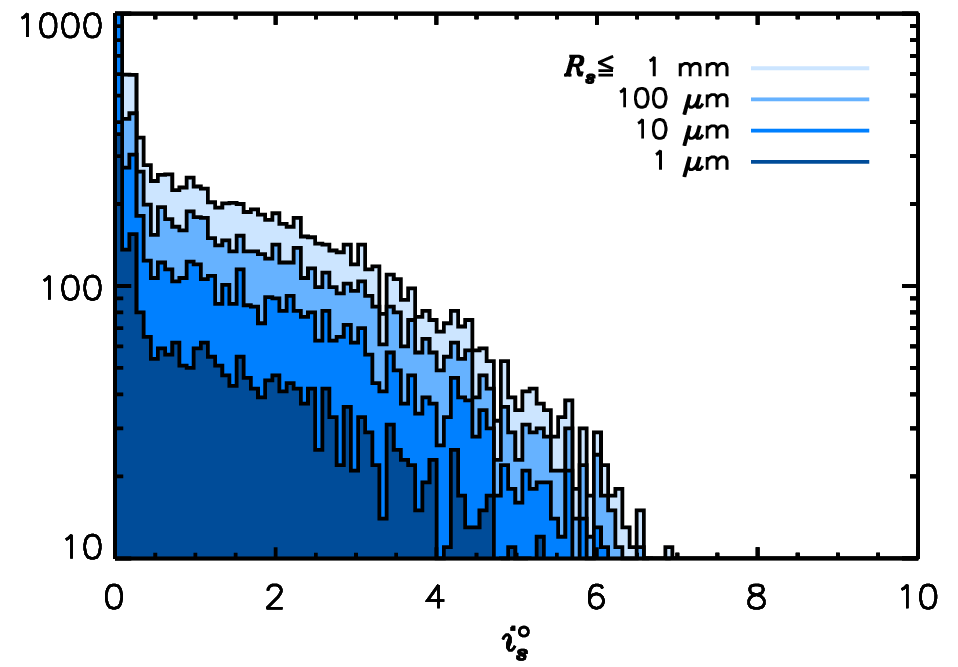}}
\caption{%
         Histograms of dust orbital properties as in \cifig{fig:hist_nop},
         but for a model that contains a giant planet.
         Top and bottom panels display distributions at a binary phase
         around apocenter and pericenter passage, respectively.
         Histograms are stacked in order of increasing dust size,
         as indicated. The enhanced concentrations around
         $2$--$2.1\,\AU$ in the left panels correspond to particles orbiting
         the planet.
         The narrow high peaks in the eccentricity distributions are generated
         by grains that move along the edges of the tidal gap opened by the planet
         (see also bottom panels of \cifig{fig:dust_3d}).
        }
\label{fig:hist_wip}
\end{figure*}

\begin{figure*}
\centering%
\resizebox{\linewidth}{!}{\includegraphics[clip]{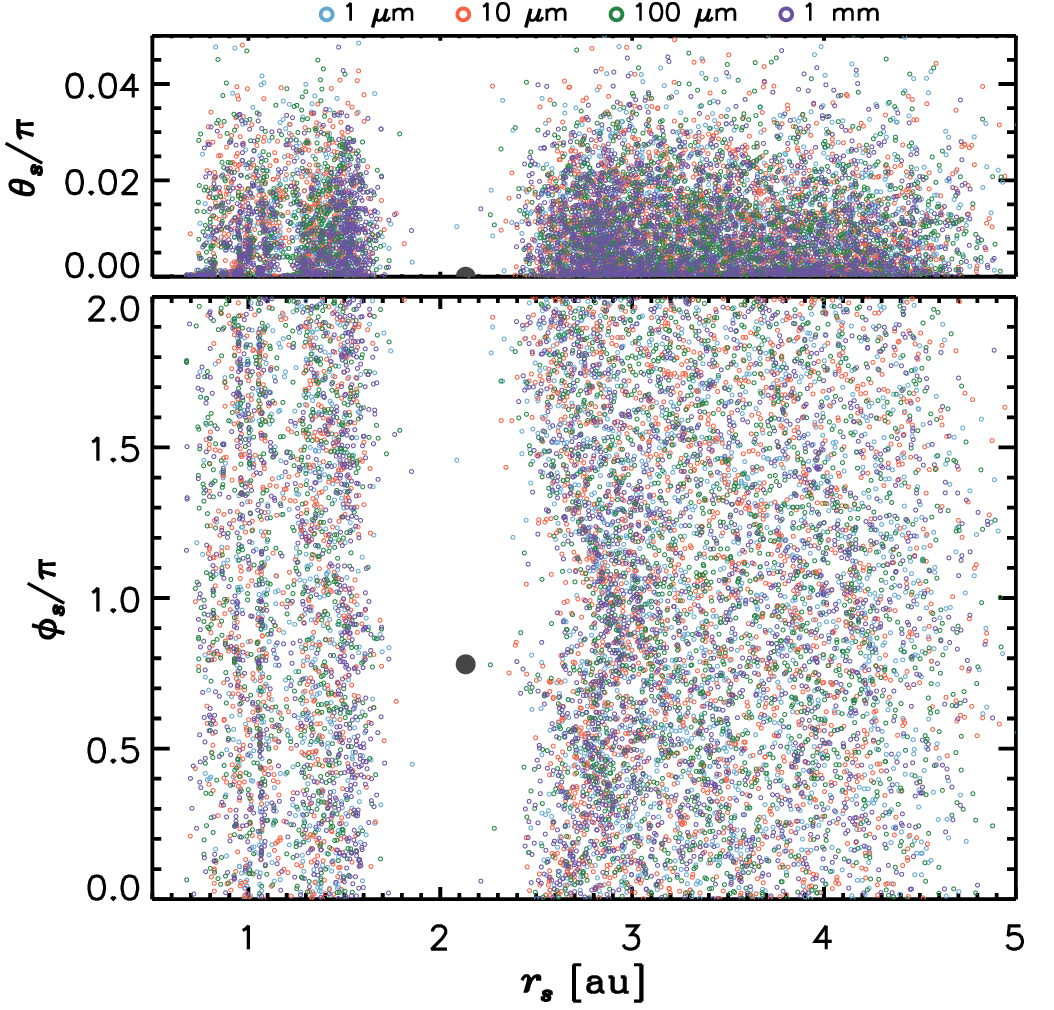}%
                          \includegraphics[clip]{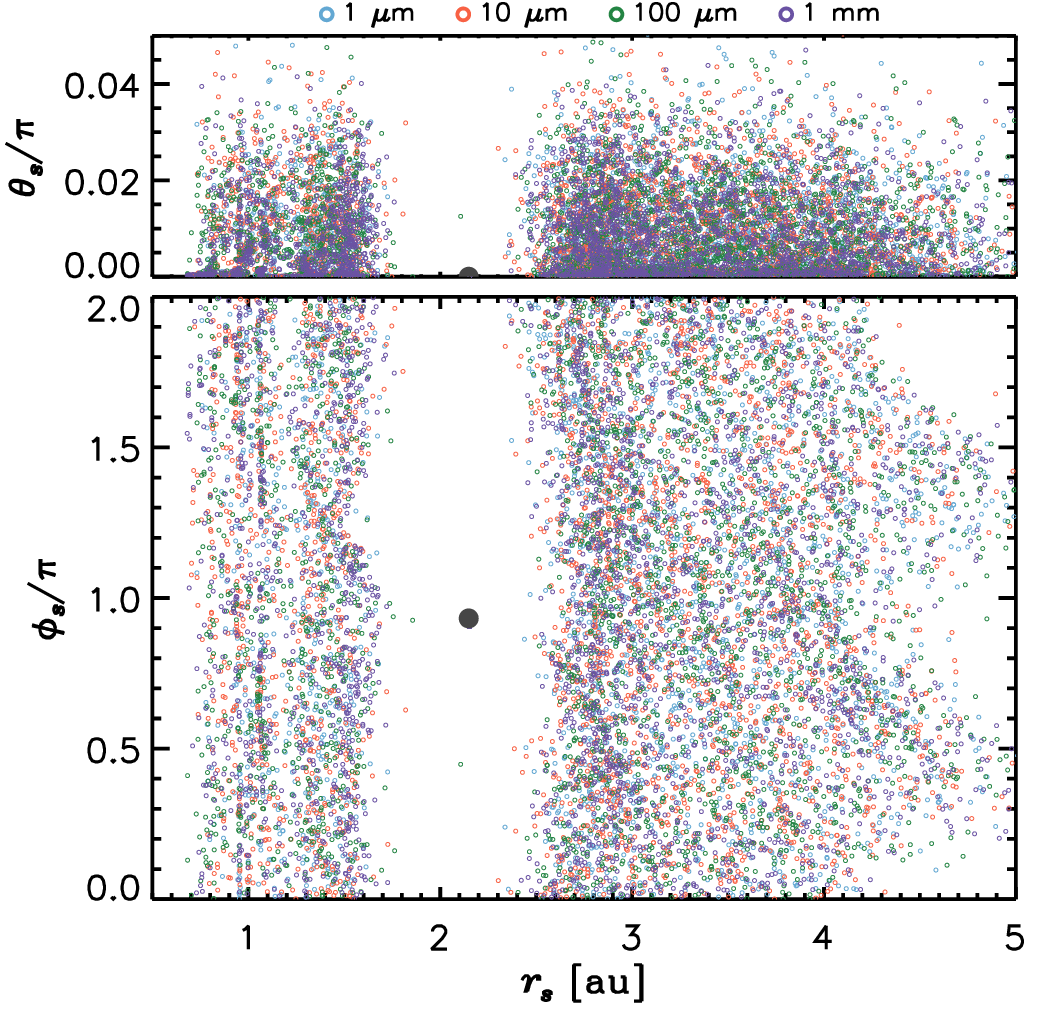}}
\caption{%
         Three-dimensional distributions of dust, for the model
         represented in \cifig{fig:hist_wip}, displayed at a binary phase
         around apocenter passage (left) and pericenter passage (right).
         The positions of the dust 
         are projected in the $r$-$\theta$ plane in the top panels 
         ($\theta_{s} = 0$ at the disk mid-plane) and $r$-$\phi$ plane. 
         Dust of different sizes are rendered by different colors, as 
         indicated in the legend. The filled gray circles represent 
         the position of the planet.
        }
\label{fig:dust_3d}
\end{figure*}

\begin{figure}
\centering%
\resizebox{0.986\linewidth}{!}{\includegraphics[clip]{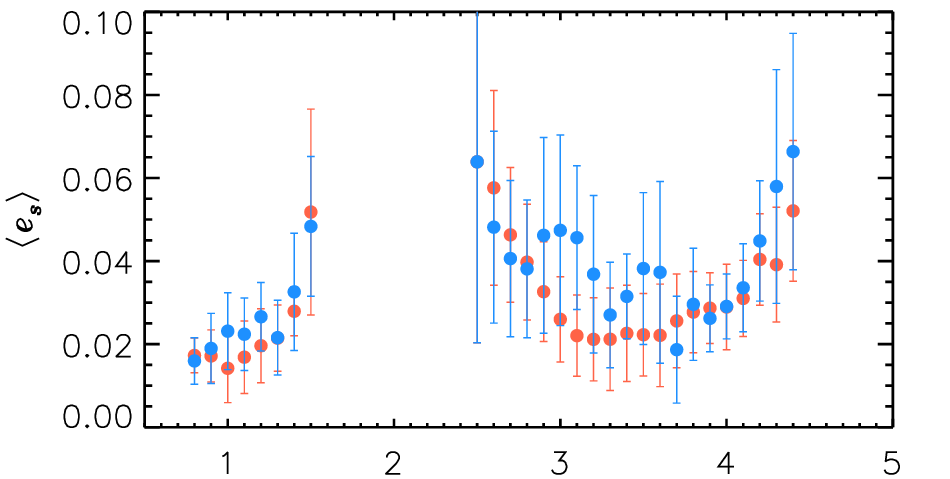}}
\resizebox{0.986\linewidth}{!}{\includegraphics[clip]{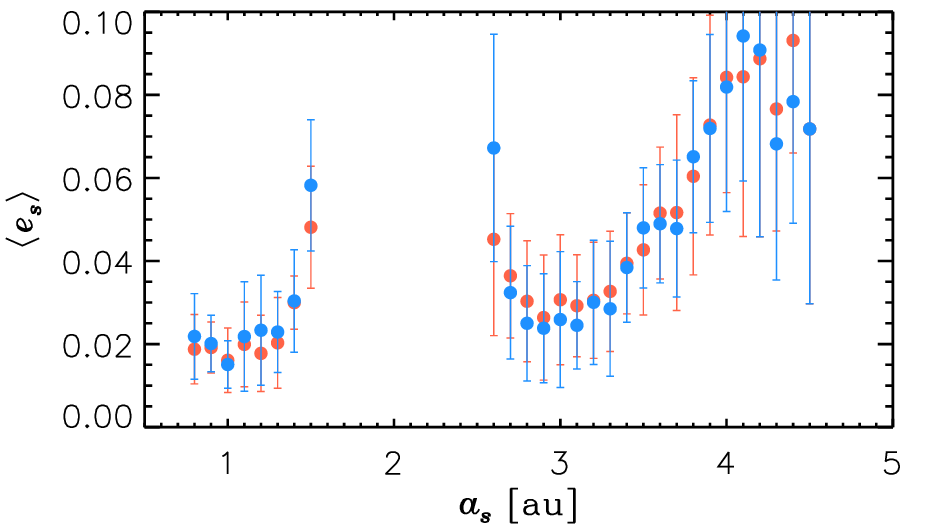}}
\caption{%
        As in \cifig{fig:es_as_nop}, but for the model with a giant
        planet displayed in \cifig{fig:hist_wip} and \cifig{fig:dust_3d}.
        }
\label{fig:es_as_wip}
\end{figure}

Silicate grains are released in the circumprimary disk after it has 
evolved for about $330\,T$, when gas has achieved
a quasi-equilibrium state (see \cisec{sec:GD}). 
The evolution of the solids is modeled for ten binary periods thereafter.
We model four size bins with radii $R_{s}=1\,\mu\mathrm{m}$,
$10\,\mu\mathrm{m}$, $100\,\mu\mathrm{m}$, and $1\,\mathrm{mm}$.
The initial distribution extends, radially, from the inner boundary of 
the computational domain out to $r=4.5\,\AU$ and, vertically, over
three pressure scale-heights $H$ from the mid-plane.

\subsection{Models without a giant planet}
\label{sec:DDnop}

The distributions of the grains' orbital semi-major axis ($a_{s}$), eccentricity
($e_{s}$), and inclination ($i_{s}$) are reported in \cifig{fig:hist_nop}, when
the secondary star is around apocenter (top) and pericenter passage
(bottom).
The histograms are stacked in order of increasing grain size, as indicated 
in the legends. The differences
at the two binary phases are marginal, indicating that the dust orbital
properties do not experience major variations during a binary period.
In particular, semi-major axes and inclinations are quite similar.
Some differences appear in the tails of the eccentricity distributions,
which extend to higher values around pericenter passage.
Statistically, however, the distributions are not distinguishable
(see also \cifig{fig:es_as_nop}), as discussed further below. 
Around apocenter, $\langle e_{s}\rangle\approx 0.04$ with a cut-off
at $e_{s}\approx 0.07$ for all grain sizes.
Around pericenter, the average eccentricity has a somewhat broader peak,
ranging from $e_{s}\approx 0.02$ to $0.06$, at all dust sizes.
These average values of eccentricity are consistent with those
estimated for the eccentricity of gas elements in the disk.

The fact that the distributions of orbital properties do not display
significant differences as a function of grain radius is mostly due
to their small stopping time, $\tau_{s}$
\citep[][]{whipple1973,weidenschilling1977b}.
Indicating with $|u|$ the relative velocity between gas and a dust
grain,
the ratio of $|u|$ to the drag acceleration defines the timescale over
which gas and dust velocities converge. In the Epstein regime of
free-molecular flow and $|u|\ll c_{s}$ \citep[e.g.,][]{liffman2000,gennaro2015}
\begin{equation}
 \tau_{s}\Omega_{\mathrm{K}}\approx
 \left(\frac{\varrho_{s}}{\rho}\right) \left(\frac{R_{s}}{H}\right).
 \label{eq:tau_s}
\end{equation}
Recall that $c_{s}=H\Omega_{\mathrm{K}}$ (the thermal velocity of the gas 
is approximated to $c_{s}$). The non-dimensional quantity
$\tau_{s}\Omega_{\mathrm{K}}$ is referred to as Stokes number,
\St. For the applied parameters, within $\approx H$ of the disk mid-plane,
$\St\lesssim 10^{-6}$ for $\mu$m-grains and $\lesssim 10^{-3}$ for mm-grains.
Since $\rho H\propto \Sigma$, $\St$ increases with $r$ (see \cifig{fig:sig_av}).
Dust dynamics is therefore well coupled to gas dynamics.
Thus, the values of the dust eccentricity quoted above are 
expected to be consistent with the estimates of the gas eccentricity, 
provided in \cisec{sec:GDnop}, for all dust sizes.
Even if the long-term values of $\langle\Sigma\rangle$ were determined
by the supply of gas from a circumbinary disk, as discussed in \cisec{sec:GDpop},
and gas density was a factor of $\sim 10$ smaller, Stokes numbers
would only be $10$ times larger. The implication is that mm-grains would
still be well coupled to the gas, and they would remain so until 
$\langle\Sigma\rangle$ declined below $\sim 10\,\mathrm{g\,cm}^{-2}$,
presumably toward the end of the circumbinary disk life.

When $\St\ll 1$, grains drift inward on a timescale
\citep[e.g.,][]{takeuchi2002,chiang2010}
\begin{equation}
 \tau_{\mathrm{drift}}\Omega_{\mathrm{K}}\sim
 \frac{1}{\St}\left(\frac{a_{s}}{H}\right)^{2},
\label{eq:tau_drift}
\end{equation}
whereas settling toward the mid-plane occurs on a typically much 
shorter timescale \citep[e.g.,][]{dullemond2004},
\begin{equation}
 \tau_{\mathrm{sett}}\Omega_{\mathrm{K}}\sim
 \frac{1}{\St}.
\label{eq:tau_sett}
\end{equation}
Note that both Eq.~(\ref{eq:tau_drift}) and (\ref{eq:tau_sett})
neglect effects of the orbital eccentricity and inclination of the grains.
The drift timescale is long enough that even mm-grains should not move
inward significantly during the modeled evolution of the disk
($\approx 790$ orbital periods at $1\,\AU$). This outcome is consistent
with the histograms of \cifig{fig:hist_nop}, left panels.
The settling timescale is short enough for some settling to occur.
Indeed, the right panels of \cifig{fig:hist_nop} indicate rapid dust
sedimentation above $\approx 2H$ of the disk mid-plane, where $\rho$
is very low and $\St$ is much larger than it is closer to the mid-plane.
The distributions of the inclination, $i_{s}$, also show that
there is an enhanced concentration of particles around the mid-plane
(see histogram peaks at $i_{s}^{\circ}< 0.5$), where the number
density is larger than the initial value by a factor of several.
These peaks appear at all particle sizes and have similar enhancements
in number density (relative to the initial values).
We performed a pair-wise comparison of the distributions of the particles'
inclination (see right panels of \cifig{fig:hist_nop}), for all sizes.
Results indicate that they are all similar, in that 
the difference in expected values (means) is compatible with zero
(i.e., it is within the standard deviations of the distributions).
This outcome is likely aided by vertical mixing induced by gas stirring 
(as shown in \cifig{fig:3d_st}).

Since the smallest, $1\,\mu\mathrm{m}$ grains are those that most closely
follow gas dynamics, they can be used as a proxy to track gas orbital
eccentricity.
\cifig{fig:es_as_nop} illustrates the eccentricity of the smallest dust,
binned over $0.1\,\AU$ intervals, as average value (dots) and standard
deviation about the mean (error bars). The red symbols refer to dust orbiting
within one scale-height $H$ from the mid-plane, whereas the blue
symbols refer to grains orbiting above $H$.
Inside $r\approx 3\,\AU$ ($r/a\approx 0.15$), the fine particles located
closer to the mid-plane have orbits somewhat more eccentric than those of
particles located farther away. At larger distances from the star, 
differences subside as orbital eccentricity increases in the low density
region of the circumprimary disk.

\subsection{Models with a giant planet}
\label{sec:DDpop}

The distributions of the dust semi-major axis, eccentricity, and
inclination, for a 3D model containing a giant planet, are plotted in
\cifig{fig:hist_wip}, when the secondary is around apocenter (top) and
pericenter passage (bottom).
The peak around $a_{s}\approx 2$--$2.1\,\AU$ is caused by particles orbiting
the planet (see Table~\ref{table:sum}).
Since the initial distribution of dust is uniform across the gap,
these grains are leftovers of that local (initial) population, most of which
is displaced from the gap region. In fact, after the region is depleted, dust
remains trapped beyond the outer edge of the gap and does not cross toward
the inner region of the disk.
The high narrow peaks in the center panels track the eccentricity of
grains that orbit along the edges of the gap. The peak's position
varies in time, possibly indicating that the average eccentricity 
of the gap edge can change in response to tidal deformation.

The spatial distributions of the dust are displayed in \cifig{fig:dust_3d},
in which particles of different size (as indicated) are projected in 
a vertical plane (top) and in the mid-planet (bottom).
In the $r$-$\phi$ plane,
the spatial distribution of the grains is mostly 
undisturbed around apocenter (right), but it is affected by the spiral
density waves of the gas at the disk edge (left), around pericenter passage.

Besides the depletion of dust around the planet orbit, the distributions in 
\cifig{fig:hist_wip} are broadly similar to the corresponding ones of
\cifig{fig:hist_nop}, indicating that the effects of the planet are mainly
confined around the orbit region.
In particular, we performed a size-wise comparison of the inclination 
distributions illustrated in the right panels of those figures, 
by means of a two-sample Kolmogorov-Smirnov test.
The results of these tests confirm that the distributions of inclination 
are not statistically distinguishable between the two models.

We also repeated the analysis in \cifig{fig:es_as_nop} for the model
with the planet, whose results are shown in \cifig{fig:es_as_wip}.
The largest differences occur around the gap edges but, otherwise,
the eccentricities as function of orbital radius display a similar 
behavior.

\section{Conclusions}
\label{sec:CC}

We use 3D hydrodynamics simulations with Lagrangian particles to analyse
the dynamics of gas and dust around the primary star of a close and eccentric
binary system, using $\gamma$ Cephei as a representative case.
We consider configurations with and without an embedded giant planet orbiting
the primary star.

Contrary to previous results \citep[e.g.,][and references therein]{jordan2021},
based on 2D simulations (and confirmed herein),
we found that 3D models predict low average orbital eccentricities 
($\lesssim 0.03$) and slow retrograde precession. We also identify the range
of pressure scale-heights in which the transition to prograde precession takes
place.
These outcomes, which apply to both gas and dust, would be consequential
for planetary assembly in the circumprimary disk since they would facilitate
accumulation of solids and planetesimal formation.

Models suggest that dust grains, in the range from $\sim \mu\mathrm{m}$ to 
$\sim \mathrm{mm}$,
remain well coupled to the gas until the gas surface density becomes very small
($\lesssim 10\,\mathrm{g\,cm}^{-2}$).
Therefore, their dynamics is dictated by that of the gas and the overall
distributions of orbital elements are statistically similar across this
size range. We also conclude that the planet does not have any significant
(large scale) impact 
on the vertical distribution of dust and only contributes locally
to the orbital eccentricity of gas and dust. 

As argued by \citetalias{marzari2025}, 3D models also suggest that a compact
circumprimary disk has a short lifetime, $\sim 10^{5}$ years, also due to
an enhanced mass transport driven by tidal interactions. 
In situ
formation of a giant planet would require a long-term supply of material from
an external source, for example, from a circumbinary disk. Indeed, the presence of
a massive planet may provide indirect proof that the binary system was once
surrounded by a disk. Under continued supply of material 
(over a few million years), the model parameters adopted herein would allow
for planet growth way beyond the mass of Jupiter, a scenario consistent with
some observations \citep{benedict2018}. 
Small dust grains would remain largely segregated beyond the outer edge 
of the gap, possibly leading to the formation of a dense dust ring.

\begin{acknowledgements}
This work benefited from conversations with Lucas Jordan, Anna Penzlin, and
Giovanni Picogna.
We thank an anonymous referee for a careful review of this paper.
Support from NASA's Research Opportunities in Space and Earth Science (ROSES)
is gratefully acknowledged.
Computational resources supporting this work were provided by the NASA 
High-End Computing Capability (HECC) Program through the NASA Advanced 
Supercomputing (NAS) Division at Ames Research Center.
\end{acknowledgements}

\end{document}